\documentclass[aps,prl,10pt,final,twocolumn,letterpaper,superscriptaddress]{revtex4-2}
\usepackage{graphicx}
\usepackage{amsmath}
\usepackage{amssymb}
\usepackage{siunitx}[=v2]
\usepackage{nicefrac}
\usepackage{color}
\usepackage{xstring}
\usepackage[normalem]{ulem}

\makeatletter
\def\@dotsep{4.5}
\makeatother

\newlength{\colwidth}
\DeclareSIUnit{\nounit}{{}}

\let\oref\ref
\renewcommand{\ref}[1]{\protect\IfSubStr{#1}{eq:}{(\oref{#1})}{\oref{#1}}}

\renewcommand{\paragraph}[1]{{\it #1.---}}

\newcommand{\horizontal}{}
\renewcommand{\deg}{^{\circ}}
\newcommand{\hdiv}{{\vec{\nabla}\!}_{\horizontal}\cdot}%
\newcommand{\hgrad}{{\vec{\nabla}\!}_{\horizontal}}%
\newcommand{\hr}{\vec{r}_{\horizontal}}
\newcommand{\hu}{\vec{u}_{\horizontal}}
\newcommand{\hPhi}{\vec{\Phi}_{\horizontal}}

\newcommand{\uav}{\overline{\vec{u}}}
\newcommand{\phiav}{\overline{\phi}}
\newcommand{\Dav}{\overline{D}}%

\newcommand{\RH}{\mathrm{RH}}

\newcommand{\hPe}{\mathrm{Pe}_{\horizontal}}

\newcommand{\DDiff}{\hPe^{-1}}
\newcommand{\orderof}{O}

\begin{document}

\title{Taylor Dispersion in Thin Liquid Films of Volatile Mixtures: A Quantitative Model for Marangoni Contraction}
\date{\today}

\author{O. Ram\'irez-Soto}
\email{olinka.ramirez@ds.mpg.de}
\author{S. Karpitschka}
\email{stefan.karpitschka@ds.mpg.de}
\affiliation{Max Planck Institute for Dynamics and Self-Organization (MPI-DS), Am Fassberg 17, 37077 G\"ottingen, Germany}

\begin{abstract}
The Marangoni contraction of sessile droplets occurs when a binary mixture of volatile liquids is placed on a high-energy surface. Although the surface is wetted completely by the mixture and its components, a quasi-stationary non-vanishing contact angle is observed. This seeming contradiction is caused by Marangoni flows that are driven by evaporative depletion of the volatile component near the edge of the droplet.
Here we show that the composition of such droplets is governed by Taylor dispersion, a consequence of diffusion and strong internal shear flow. We demonstrate that Taylor dispersion naturally arises in a self-consistent long wave expansion for volatile liquid mixtures. Coupled to diffusion limited evaporation, this model quantitatively reproduces not only the apparent shape of Marangoni-contracted droplets, but also their internal flows. 
\end{abstract}

\maketitle

\begin{figure}[t!]%
	\centering\includegraphics[height=86mm]{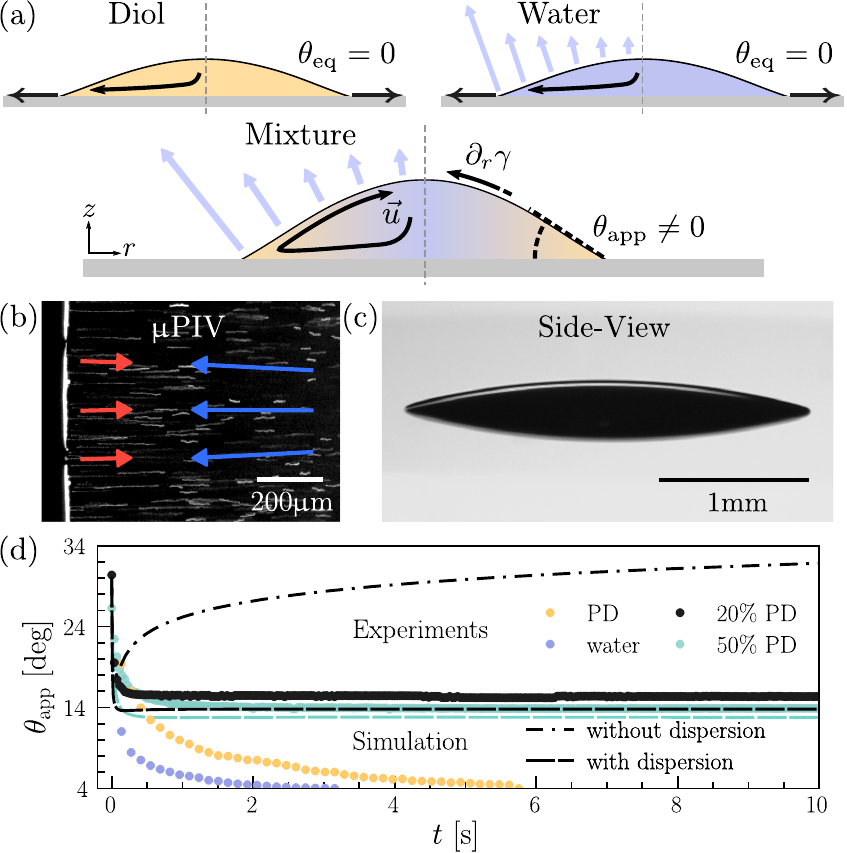}%
	\caption{\label{fig:FIG1}(a) Pure droplets of water or diols spread on hydrophilic glass, but their mixtures display Marangoni contraction: Evaporative enrichment drives a Marangoni flow which compensates the capillarity-driven spreading flow. (b) Streak image of fluorescent particles inside a drop, in a plane parallel to the substrate. Close to the contact line, the flow is directed into the droplet. (c) Side aspect of the same drop. (d) Apparent contact angle vs. time  for a relative humidity $\sim 35\%$: experiments (symbols) with pure liquids (blue \& yellow) and mixtures (black: 20\% PD; mint: 50\% PD), compared to long-wave simulations of the mixtures (lines), with and without Taylor dispersion (dashed and dash-dotted lines, respectively). Simulations are in close agreement with experimental data only when dispersion is taken into account.}%
\end{figure}

Wetting and dewetting of volatile liquid mixtures on solid surfaces is abundant in natural phenomena and technological applications~\cite{Lohse:NRP2020,Brutin:ChemSocRev2018,Snoeijer:ARFM2013,Bonn:RMP2009,Smith:L2018}. Many examples are found in everyday life situations, for instance, biological fluids such as blood~\cite{Brutin:JFM2010} and tears~\cite{Traipe-Castro:BR2014}, inks for inkjet printing~\cite{DeGans:L2004,Park:L2006}, and paints for artistic techniques~\cite{Zenit:PRF2019,GiorgiuttiDauphine:JAP2016}.
Marangoni contraction is a prime example that gained significant attention recently, not least motivated by its applications in printing and semiconductor processing~\cite{Cira:N2015,Karpitschka:L2017,Benusiglio:SM2018,Sadafi:L2019,
Malinowski:SA2020,Williams:JFM2020,Hack:L2021,Shiri:PRL2021}:
Volatile liquids seemingly dewet from high energy surfaces over which they spread if evaporation was suppressed, see Fig.~\ref{fig:FIG1}~(a, c).
Evaporation causes compositional~\cite{Cira:N2015,Karpitschka:L2017,Benusiglio:SM2018,Williams:JFM2020,Hack:L2021} or thermal~\cite{Shiri:PRL2021} gradients, inducing an inward Marangoni flow which contracts the droplet.
The opposite case may lead to Marangoni spreading and contact-line instabilities~\cite{Darhuber:PoF2003,Gotkis:PRL2006,Wodlei:NC2018}.
The dynamics of contact lines is a multi-scale problem which involves both macroscopic hydrodynamics and molecular interactions~\cite{Bonn:RMP2009,Oron:RMP1997}.
Thus, while multi-component liquids with pinned contact lines are understood quite well~\cite{Zhang:PoF2011,Christy:PRL2011,Soulie:PCCP2015,Marin:PRF2019,Rossi:PRE2019,
Karapetsas:L2016,Diddens:JCP2017,Diddens:JFM2017,Li:PRL2019,Gaalen:JCIS2021,Pahlavan:PRL2021}, moving contact lines challenge both experimentalists and theoreticians.

Dimensional reduction in the limit of long waves is a powerful tool to analyze moving contact lines~\cite{Oron:RMP1997,Craster:RMP2009,Gaalen:JCIS2021,Thiele:PoF2012,
Thiele:PRF2016,Xu:JPCM2015,Gaalen:JCIS2021}:
The evolution of the local liquid height is derived from the net flux, treating the short (vertical) axis fully implicitly.
For mixtures, however, vertical compositional gradients are unavoidably generated by shearing any horizontal gradients.
This impedes the use of dimensional reduction, unless taking the limit of infinitely fast diffusion along the short axis.
All existing lubrication models are formulated in this limit~\cite{Moshinskii:FD2004,Ajdari:AC2006,Mukahal:PRSA2017,Vilquin:a2020}.
Commonly, however, the time scale of diffusion is finite and, in combination with shear flow, leads to strong dispersion.
This so-called Taylor-Aris dispersion~\cite{Taylor:PRSA1953,Aris:PRSA1956} has important consequences in many natural~\cite{Chakrabarti:PoF2020} and technological scenarios~\cite{Darhuber:PRSA2004}.
To date it remains unclear whether shear dispersion is consistent with a long-wave expansion, so no expression for the effective dispersion in general thin-film flows is available in the literature.

Here we show that small vertical compositional gradients are in agreement with the usual assumptions in a long-wave expansion. We provide a general expression for the effect of shear dispersion in thin liquid films, thus enabling lubrication theory to be consistently applied to bulk liquid mixtures.
Our model is in quantitative agreement with experimental observations of Marangoni-contracted droplets.
We expect our analysis to be relevant well beyond droplet studies, as it offers a general route for implementing the effect of advected bulk fields in dimensional reduction problems.


\begin{figure}[t]%
	\centering\includegraphics{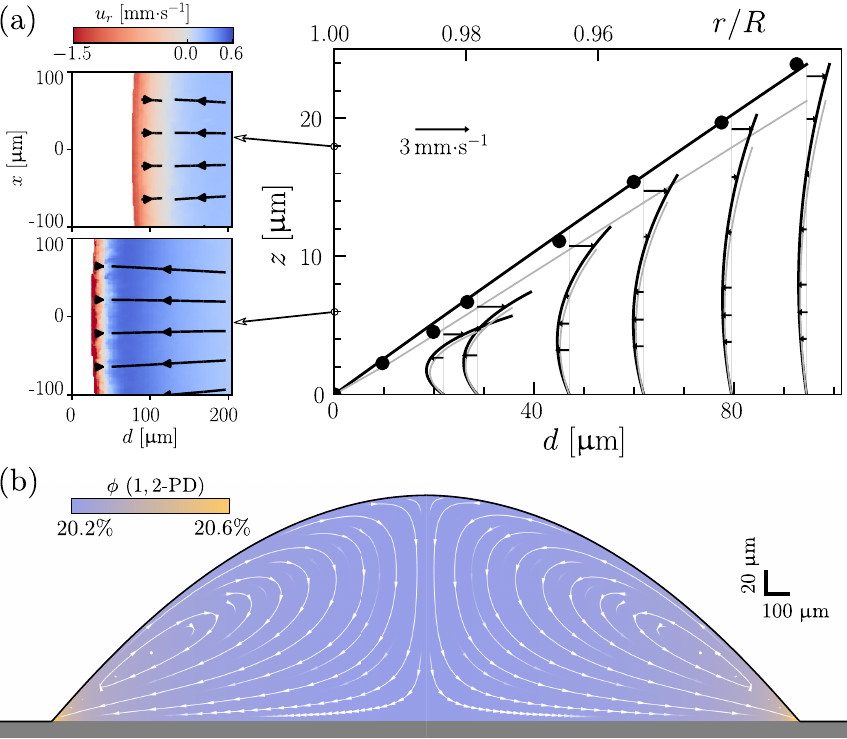}%
	\caption{\label{fig:FIG2}Flows in a Marangoni-contracted droplet ($\phi=0.20$ $1,2$-PD, $\mathrm{RH}=33\%$). (a)~Cross-sectional view: free surface (black discs \& linear fit), horizontal velocity (black arrows) and parabolic fits (black lines) from experiments, compared to simulation results including shear dispersion (gray). Insets: experimental velocities and stream lines in two horizontal planes ($z= \SI{18}{\micro \meter}$ and $\SI{6}{\micro \meter}$). (b)~Simulation snapshot ($t=\SI{8.0}{\second}$) with stream lines (white) and composition (color code).}%
\end{figure}%

\paragraph{Experiments}%
We measured the apparent shape and the internal flows of Marangoni-contracted droplets inside an atmospheric control chamber (size $\sim \SI{10 x 10 x 10}{\cm}$), at room temperature. The droplets were composed of mixtures of water (``Milli-Q'', resistivity $\SI{18}{\mega\ohm\cm}$) and a carbon diol (Sigma Aldrich, $\geq 98\%$). Piranha cleaned microscopy glass coverslips ($\SI{170}{\micro\metre}$ thick) were used as substrates (see Supplemental Material for details~\cite{supplement}\nocite{
Leonard:IJNME1990,Cohen:CP1996}%
).
The humidity was set by continuously injecting a well-defined mixture of dry and moist nitrogen behind gas-permeable membranes at the side-walls of the chamber. The droplets had initial volumes of $0.5$ to $\SI{1}{\micro\liter}$.
Micro particle image velocimetry (\SI{}{\micro\nounit}PIV, Fig.~\ref{fig:FIG1}~(b)) was performed with an inverted fluorescence microscope and a high-aperture water-immersion objective (20x, numerical aperture (NA) 0.95) to allow for diffraction limited imaging in the bulk droplet.
Polystyrene microspheres (Thermo Fisher Scientific F8809, diameter $\SI{200}{\nano\meter}$) were used as flow tracers, with a mass fraction of $\SI{7.8e-5}{}$ of the particle stock solution in the binary mixture.
Images of the particles were captured with a high-speed camera at $600$ to $\SI{1000}{FPS}$, quickly switching between $z$-planes by automating the focus system of the microscope. Simultaneous side-view imaging of the drop was performed with a telecentric lens (Fig.~\ref{fig:FIG1}~(c)).

Fig.~\ref{fig:FIG1}~(d) shows the evolution of the apparent contact angle $\theta_{app}$ of spreading drops of pure water (diol mass fraction $\phi=0$), pure 1,2-propanediol (PD, $\phi=1$) and of their mixtures. Pure liquids spread into complete wetting. In contrast, the binary mixtures reach a stationary non-equilibrium apparent contact angle $\theta_{app}>0$ shortly after deposition. The drops stay in this contracted state for several minutes. This wetting behavior has been described previously~\cite{Cira:N2015,Benusiglio:SM2018,Karpitschka:L2017}, showing that $\theta_{eq}$ depends on $\phi$ and the ambient relative humidity $\RH$.

Fig.~\ref{fig:FIG2}~(a) shows the velocity field inside the droplet, as determined by \SI{}{\micro\nounit}PIV. The arrows indicate the velocities that have been measured in different $z$-planes. The insets show dense velocity fields for two exemplary $z$-planes.
Close to the free surface, the flow is directed inward, precisely balanced by an outward flow close to the substrate, leading to a quasi-stationary shape.

\begin{figure}[t]%
  \centering
      \includegraphics{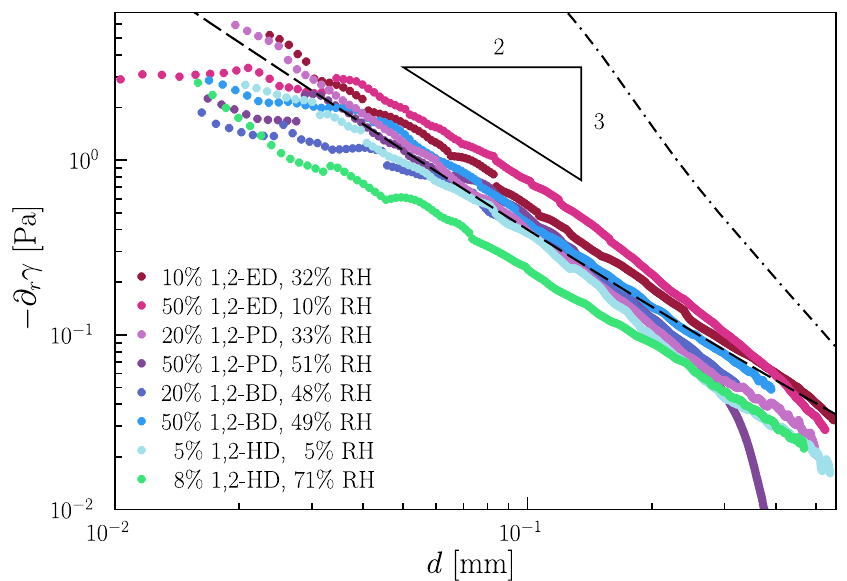}%
	\caption{\label{fig:FIG3}Surface tension gradient vs. distance to the contact line. Experimental data for various compositions and humidities (dots) follows a $\sim d^{\nicefrac{-3}{2}}$ power law. Simulations ($\phi=0.20$ $1,2$-PD, $\mathrm{RH}=33\%$, $t=\SI{8.0}{\second}$) with Taylor-Aris dispersion (dashed line) are in quantitative agreement with experiments, in contrast to the results obtained without Taylor-Aris dispersion (dash-dotted line). 
}
\end{figure}

The surface tension gradient can be derived from the tangential stress boundary condition, using the measured shear rate and the viscosity $\eta(\phi)$ from the literature \cite{Moosavi:JCED2017,George:JCED2003,Jarosiewicz:ZNA}.
Fig.~\ref{fig:FIG3} shows the experimentally derived surface tension gradient as a function of the distance $d$ to the contact line, for various diols, compositions, and relative humidities. All curves follow a power law $\partial_r \gamma \sim d^{\nicefrac{-3}{2}}$.
Despite significant differences in composition, surface activity, and ambient humidities, the curves nearly collapse in physical units.

\paragraph{Lubrication theory}%
We consider the general case of a thin liquid film of a mixture on a flat solid surface. 
The film covers the entire substrate, with a continuous transition between the macroscopic droplet and a microscopically thin precursor surrounding it.
The latter reflects the adsorption equilibrium of vapor molecules in the atmosphere around the droplet~\cite{Eggers:PoF2010}. For water above the dew point, adsorption layers on hydrophilic surfaces are typically on the order of a few $\SI{}{nm}$~\cite{Verdaguer:L2007,Barnette:PCCP2008}.
The free surface is described by $h(\hr)$, where $\hr$ is the location in the substrate plane (see Fig.~\ref{fig:FIG1}~(a)).
Incompressible Stokes flow without body forces is governed by
\begin{align}
	\label{eq:stokes}
  \eta\vec{\nabla}^2 \vec{u} &= \vec{\nabla}p\text{,}\\
	\label{eq:ucontinuity}
	\vec{\nabla}\cdot\vec{u} &= 0
\end{align}
where $\vec{u}$ is the fluid velocity, $\eta$ is the dynamic viscosity of the fluid, and $p$ is the fluid pressure.
The evolution of the solute field $\phi$ is given by
\begin{equation}
	\label{eq:phicontinuity}
	\partial_t \phi = \vec{\nabla}\cdot\left( D\,\vec{\nabla}\phi - \vec{u}\,\phi \right)\text{,}
\end{equation}
with $t$ as time and $D$ as the diffusion coefficient of the solute. For simplicity we limit the following analysis to isothermal, isochoric, and isoviscous cases. We consider a no-slip and no-flux boundary condition at $z=0$, kinematic and stress boundary conditions at $z=h$, and a Stefan-type boundary condition that links composition and evaporation (see Supplemental Material for details~\cite{supplement}).

To derive evolution equations in terms of vertically averaged quantities, we take the limit of long waves, where the characteristic horizontal scale, $r_0$, shall be much larger than the characteristic vertical scale, $h_0$: $\epsilon_{h} = h_0 / r_0 \ll 1$. For sessile droplets, $r_0$ and $h_0$ are conveniently chosen as the footprint radius and the apical height of the droplet, respectively.
We define the vertically averaged velocity $\uav$, the total hydrodynamic flux $\vec{\Phi}$, the vertically averaged composition $\phiav$, and the effective solute height $\Psi$ through
\begin{equation}
	\label{eq:Phidef}
	\vec{\Phi} = \int_0^h\!\! dz\, \vec{u}\text{,} = \uav\,h\text{,}\quad\quad
	\Psi = \int_0^h\!\! dz\, \phi\text{,} = \phiav\,h\text{,}
\end{equation}
and the deviations from the average by
\begin{equation}
	\vec{u} = \uav + \delta\vec{u}\text{,}\quad\quad
	\label{eq:phisplit}
	 \phi = \phiav + \delta\phi\text{.}
\end{equation}
We scale all horizontal coordinates as $\hr = r_0\, \hr'$, and all vertical coordinates as $z = \epsilon_h\, r_0 z'$.
Velocities and time are scaled with $u_0$ and $t_0$, the characteristic velocity and time scales of the problem, respectively. Below, $u_0$ and $t_0$ will be identified with the natural scales that arise in the evolution equations. Surface tension is scaled as $\gamma = \gamma_0\,\gamma'$, where $\gamma_0 = \gamma(\phi=0)$, the surface tension of the pure solvent. Diffusivity is treated similarly, $D = D_0\, D'$ with $D_0 = D(\phi=0)$. We scale $j = j_0\, j'$ for the evaporation rate, where $j$ and $j_{0}$ are determined according to Ref.~\cite{Eggers:PoF2010} for the diffusion limited regime.

Whether $\phiav$, the vertically averaged composition, is sufficient to describe the compositional evolution of the film, depends on the magnitude of the residual field $\delta\phi$. Thus we scale $\delta\phi = \epsilon_{\phi}\, \delta\phi'$, deriving $\epsilon_{\phi}$ from the governing equation for $\delta\phi$. In the following we will omit the primes for readability and work exclusively with scaled quantities.


The derivation of the evolution equation for the film height follows the standard procedure described in the reviews~\cite{Oron:RMP1997,Craster:RMP2009}. One obtains
\begin{equation}
  \partial_t h	= -\hdiv\hPhi - \mathcal{E}\, j\text{,}
	\label{eq:hevol}
\end{equation}
from integrating~\ref{eq:ucontinuity} along $z$, where $\mathcal{E} = j_0\,\eta/(\epsilon_h^4\, \gamma_0)$. Note that~\ref{eq:hevol} does not involve any approximation. The long wave expansion is used only in the expressions for the evaporation rate $j$ (see Ref.~\cite{Eggers:PoF2010}) and the horizontal hydrodynamic flux~\cite{Matar:PoF2002} 
\begin{equation}
	\label{eq:Phi}
	\hPhi= \vec{\Phi}_C + \vec{\Phi}_M = 
	  - \frac{h^3}{3}\hgrad p
	  + \frac{1}{\epsilon_h^2}\frac{h^2}{2} \hgrad\gamma
	  + \orderof(\epsilon_h^2) \text{.}
\end{equation}
Here we identified the natural velocity scale $u_0 = \epsilon_h^3\,\gamma_0/\eta$, the capillary velocity for thin films, to cancel the material properties from Eq.~\ref{eq:Phi}. For a typical 1,2-propanediol/water droplet with $\phi\sim0.2$ (i.e., $\gamma\sim\SI{56}{mN/m}$ and $\eta\sim\SI{2}{\milli\pascal\second}$~\cite{Karpitschka:L2010}), $r_0\sim \SI{1.5}{mm}$, and $h_0\sim\SI{0.15}{mm}$ one obtains $u_0\sim \SI{28}{mm/s}$, much larger than the maximum experimentally observed velocities $\sim\SI{2}{mm/s}$. This is a common observation in wetting problems, where the capillary number typically remains small. The natural pressure- and time scales are $p_0 = \epsilon_h\,\gamma_0/r_0$ and $t_0 = r_0\,\eta / (\epsilon_h^3 \gamma_0)$, respectively. The pressure $p = -\gamma \hgrad^2 h + \Pi(h)$ contains capillary and surface (disjoining) forces that stabilize the precursor film.

The capillary- ($\vec{\Phi}_C$) and Marangoni ($\vec{\Phi}_M$) fluxes are associated with Poiseuille- and Couette-type velocity profiles, respectively~\cite{supplement}.
These will shear any horizontal compositional gradient, such that a vertical gradient arises naturally. Combined with molecular diffusion from Eq.~\ref{eq:phicontinuity}, this leads to Taylor-Aris dispersion~\cite{Taylor:PRSA1953,Aris:PRSA1956}. In addition, the Stefan boundary condition for evaporation at the free surface requires a vertical compositional gradient~\cite{Karpitschka:CES2015,Hennessy:JCIS2017}. It is commonly accepted that vertical compositional gradients are beyond the limit of the lubrication expansion~\cite{Matar:PoF2002,Oron:PRE2004,Shklyaev:PoF2007,Shklyaev:PRE2011,
      Thiele:EPJST2011,Thiele:PRL2013,Xu:JPCM2015,Thiele:PRF2016}.
We challenge this paradigm, identifying three regimes, depending on aspect ratio and P\'eclet number:
i.~A regime of faint vertical compositional gradients where previous long-wave models hold~\cite{Matar:PoF2002};
ii.~An intermediate regime of small but not negligible vertical gradients for which we derive a previously unknown  evolution equation for $\phiav$ including Taylor-Aris dispersion;
and iii.~A regime of large vertical gradients where the full advection-diffusion problem has to be solved~\cite{Matar:PoF2002}.

Inserting~\ref{eq:phisplit} into~\ref{eq:phicontinuity} and integrating over the film height gives
\begin{align}
	      h\,\partial_t \phiav
	&=\quad  \DDiff\,\hdiv \left(   \Dav\,h\,\hgrad\phiav
	                                  + \epsilon_{\phi}\int_0^h\!\! dz\, D\,\hgrad\delta\phi
														 \right)\nonumber\\
	&\quad - \vec{\Phi}\cdot\hgrad\phiav
	       - \epsilon_{\phi}\,\hdiv\int_0^h\!\! dz\,\delta\phi\,\hu
	       + \mathcal{E}\,\phi(h)\, j\text{,}
\label{eq:phisplitint}
\end{align}
where $\hPe = u_0\, r_0 / D_0$ is the P\'eclet number and $\Dav$
is the vertically averaged diffusion coefficient
(see Supplemental Material~\cite{supplement} for a detailed derivation).
In contrast to Eq.~\ref{eq:hevol}, terms with non-averaged quantities remain. These terms scale as $\sim\epsilon_{\phi}$, while the next-order terms in~\ref{eq:hevol} with~\ref{eq:Phi} scale as $\sim\epsilon_{h}^2$. Thus, the magnitude of $\epsilon_{\phi}$ relative to $\epsilon_{h}$ determines which terms in~\ref{eq:phisplitint} should be kept.

Limit i: $\epsilon_{\phi}\,\delta\phi\lesssim\epsilon_h^2$. We may ignore all terms $\sim\epsilon_{\phi}$ and recover the previously known evolution equation~\cite{Matar:PoF2002,Jensen:PoFA1993,Jensen:CES1994}:
\begin{equation}
	h\partial_t \phiav = 
	    \DDiff\,\hdiv\Dav\,h\,\hgrad\phiav
	  - \hPhi\cdot\hgrad\phiav
	  + \mathcal{E}\,\phiav\, j
	  + \orderof(\epsilon_h^2)
	   \text{.}
\label{eq:phiavevol:limI}
\end{equation}

Limit iii: $\epsilon_{\phi}\,\delta\phi\sim 1$. No simplified evolution equation for vertically averaged quantities can be derived, and the full problem has to be solved~\cite{Matar:PoF2002,Jensen:CES1994}.

Limit ii: $\epsilon_{\phi}\,\delta\phi\sim\epsilon_h$. In this case, terms up to $\sim\epsilon_{\phi}$ have to be retained.
We consider a convection dominated problem i.e., $\hPe\gg 1$ and $\mathcal{E}\ll 1$. Far below the boiling point, and with typical $D_0\sim \SI{e-9}{m^2/s}$ and $u_0\sim \SI{e-3}{m/s}$, this holds for most sessile nano- to microliter droplets with small contact angles.
Eq.~\ref{eq:phisplitint} simplifies further~\cite{supplement}:
\begin{align}
	      h\,\partial_t \phiav
	&=\quad   \DDiff\,\hdiv \Dav\,h\,\hgrad\phiav
	        - \vec{\Phi}\cdot\hgrad\phiav\nonumber\\
	&\quad  - \epsilon_{\phi}\,\hdiv\int_0^h\!\! dz\,\delta\phi\,\hu
	        + \mathcal{E}\,\phiav\, j
	        + \orderof(\epsilon_{\phi}^2)\text{.}
\label{eq:phisplitintsimp}
\end{align}
The remaining term with $\delta\phi$ scales as $\epsilon_{\phi}$. Thus a governing equation for $\delta\phi$ can be truncated to $\orderof(1)$. Inserting~\ref{eq:phisplit} into~\ref{eq:phicontinuity}, multiplying with $h$, and subtracting~\ref{eq:phisplitintsimp} gives the leading order governing equation for $\delta\phi$~\cite{supplement}:
\begin{equation}
	  \Dav\,\partial_z^2 \delta\phi
	=   \mathcal{E}\,\phiav\, j/h 
		+\delta\hu\cdot\hgrad\phiav
		+\orderof(\epsilon_{\phi})\text{,}
\label{eq:deltaphigov}
\end{equation}
where we recover the natural scale of the residual field, $\epsilon_{\phi} = \epsilon_h^2\hPe$~\cite{Jensen:PoFA1993,Jensen:CES1994}. $\epsilon_{\phi}$ is equivalent to a P\'eclet number for the characteristic vertical length- and velocity scales, $\epsilon_h\, r_0$ and $\epsilon_h\, u_0$, respectively.
Eq.~\ref{eq:deltaphigov} defines the advection-diffusion problem of the residual field in the co-moving frame of the mean flow: Diffusion along $z$ balances the shearing due to the horizontal flow, and the residual field is stationary at leading order.

The requirements for limit ii became apparent now: The problem must be convection dominated in the horizontal direction ($\hPe\gg 1$), but diffusion dominated in the vertical direction, i.e., the residual field must remain small: $\epsilon_h^2\,\hPe\, \delta\phi\ll 1$. This is similar to the classical treatment of pipe flow by Taylor and Aris~\cite{Taylor:PRSA1953,Aris:PRSA1956}, but here the velocity field and the film height, and thus $\delta\phi$ vary in space.
By scaling $z$ in Eq.~\ref{eq:deltaphigov} with the local film height $h$, it becomes apparent that $\delta\phi\sim h^2\,\delta\vec{u}\,\hgrad\phiav$ if shear ($\delta\vec{u}$) dominates.
For Marangoni-contracted droplets, we find $\delta\phi\ll 1$ everywhere:
$\hgrad\phiav$ and $\delta\vec{u}$, which are caused by evaporative enrichment and Marangoni convection, are strong only near the edge of the droplet where the height is small (see below for a quantitative estimate).

Eq.~\ref{eq:deltaphigov} can be integrated, and the resulting expressions for $\delta\phi$ and the integral in~\ref{eq:phisplitintsimp} can be found in the Supplemental Material~\cite{supplement}.
In cases of axial or translational symmetry and slow evaporation ($\mathcal{E}\,j/h\ll\delta\vec{u}$), which holds for our experiments, Eq.~\ref{eq:phisplitintsimp} reduces to Eq.~\ref{eq:phiavevol:limI} with $\Dav$ replaced by $\Dav_{\mathrm{eff}}$ to account for Taylor-Aris dispersion:
\begin{equation}
\label{eq:Deff}
	\Dav_{\mathrm{eff}} = \Dav + \frac{\epsilon_h^2\hPe^2}{\Dav}
		\left(\frac{2\vec{\Phi}_C^2}{105} + \frac{\vec{\Phi}_C\cdot\vec{\Phi}_M}{20} + \frac{\vec{\Phi}_M^2}{30}\right)\text{.}
\end{equation}


\paragraph{Numerical simulations}%
We implemented the evolution equations~\ref{eq:hevol} and~\ref{eq:phiavevol:limI} with the flux~\ref{eq:Phi} and the effective diffusivity~\ref{eq:Deff} in an axisymmetric finite volume scheme with convergent numerical mobilities after Refs.~\cite{Diez:PRE2000,Gruen2002}, and diffusion limited evaporation according to Refs.~\cite{Eggers:PoF2010,Karpitschka:L2017}.
We used accurate material properties $\gamma(\phi)$, $\eta(\phi)$, found in the literature~\cite{Moosavi:JCED2017,George:JCED2003,Jarosiewicz:ZNA,Karpitschka:L2010}, and assumed $D\sim\eta(\phi)^{-1}$ in accordance with the Stokes-Einstein relation.
Simulations were initiated with a droplet of $\sim\SI{0.7}{\micro\liter}$ volume and $\sim30\deg$ apparent contact angle, on top of a precursor in equilibrium with the vapor field of the droplet. See~\cite{supplement} for details.

\begin{figure}[t]%
  \centering
      \includegraphics{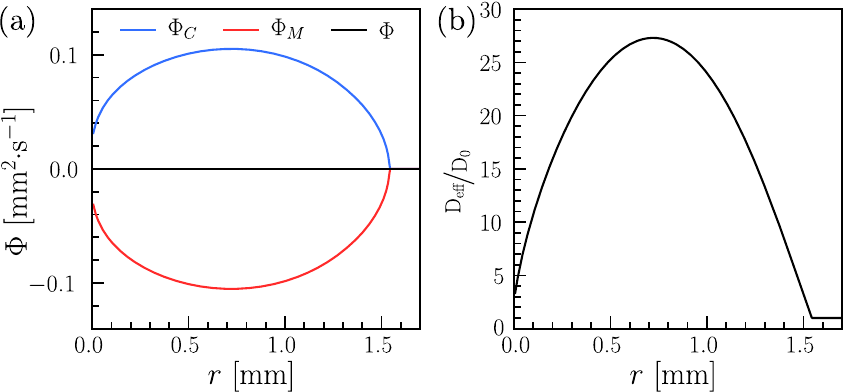}
	\caption{\label{fig:FIG4}Simulated fluxes (a) and effective diffusivity (b) as a function of $r$ ($\phi=0.20$ $1,2$-PD, $\mathrm{RH}=33\%$, $t=\SI{8.0}{\second}$). (a) Capillary (blue) and Marangoni (red) fluxes balance each other, resulting in a total flux $\Phi\sim 0$ (black). Although the net hydrodynamic transport vanishes, the different flow profiles of the two fluxes lead to strong shear dispersion. (b) Effective diffusivity (Eq.~\ref{eq:Deff}) scaled by the molecular diffusivity. Shear dispersion increases the effective diffusivity $\sim 27$-fold as compared to molecular diffusion alone.
}
\end{figure}%

Fig.~\ref{fig:FIG1}~(d) compares the apparent contact angle observed in simulations with (dashed) and without (dash-dotted) Taylor-Aris dispersion. A near-quantitative agreement is observed for stationary contraction only if dispersion is taken into account.
The remaining quantitative deviation is much smaller than the mismatch for simulations without dispersion and can be attributed to uncertainties in the material parameters, most importantly, the molecular diffusivity.
We deliberately refrain from any parameter fittings.
Fig.~\ref{fig:FIG2}~(b) shows stream lines and composition for a contracted droplet.
The observed difference in composition between the center and edge is merely $0.4\%$. The most striking feature of the simulations is a quantitative reproduction of the experimentally observed velocities (Fig.~\ref{fig:FIG2}~a, gray, simulation, vs. black, experiments) and surface tension gradients (Fig.~\ref{fig:FIG3}) for the case with Taylor-Aris dispersion.

The origin and importance of Taylor-Aris dispersion are highlighted in Fig.~\ref{fig:FIG4}. Panel (a) shows the strong but compensating capillary and Marangoni fluxes in the droplet.
Although the net hydrodynamic flux vanishes, the different velocity profiles of capillary and Marangoni fluxes lead to a convection roll inside the droplet, and thus strong shear dispersion. This leads to an effective diffusivity (including shear dispersion and molecular diffusion, Fig.~\ref{fig:FIG4}~(b)) which scales quadratically with the individual flux components (Eq.~\ref{eq:Deff}), meaning strong dispersion in regions where the fluxes are large. Here, the effective diffusivity reaches $\Dav_{\mathrm{eff}} \sim 27\Dav$ around $r\sim R/2$. Thus, Taylor-Aris dispersion becomes the governing phenomenon for the solute distribution in Marangoni-contracted droplets.

Inserting again our characteristic quantities ($r_0\sim\SI{1.5}{mm}$, $h_0\sim\SI{0.15}{mm}$, $\phi\sim0.2$, $\gamma\sim\SI{56}{mN/m}$, and $\eta\sim\SI{2}{\milli\pascal\second}$), we obtain $\hPe\sim 4\cdot 10^4$ and $\epsilon_h^2\,\hPe\sim 4\cdot 10^2$.
Thus the applicability of Taylor dispersion for our droplets depends on the magnitude of the scaled $\delta\phi\sim h^2\, \delta u\, \partial_r \phiav$.
The largest measured velocities were $\sim\SI{2}{mm/s}\sim 0.1 u_0$, close to the contact line where the scaled height $h\sim 0.1$ (see Fig.~\ref{fig:FIG2}). The gradient of the mean composition in that region can be estimated from Fig.~\ref{fig:FIG3} as $\partial_r\phiav\sim \left(\frac{\partial\gamma}{\partial\phi}\right)^{-1}\, \partial_r\gamma\sim 0.05$. Thus, the deviation from the mean composition is only 
$\epsilon_h^2\,\hPe\, \delta\phi\sim 2\cdot 10^{-2}$, which justifies the approximations.

\paragraph{Conclusion}%
We measured the internal flow fields of Marangoni-contracted drops and derived the surface tension gradient, which follows a power law $\sim d^{\nicefrac{-3}{2}}$. Through a systematic long wave expansion for free-surface films of mixtures, we could extend the widely used thin-film evolution equations to the case of bulk mixtures subject to Taylor-Aris dispersion.
For Marangoni-contracted drops, Taylor-Aris dispersion governs the composition, and our model is in quantitative agreement with the experimental findings.
The theoretical analysis enables lubrication theory to be used for the very general case of advection-dominated thin free-surface films with advected bulk fields like temperature or composition.


\emph{Note added.} Recently, we became aware of another study that includes Taylor dispersion in the description of Marangoni-contracted droplets~\cite{Charlier:JFM2022}.

\begin{acknowledgments}
\emph{Acknowledgments.} We acknowledge financial support from the Max Planck -- University of Twente Center for Complex Fluid Dynamics. S.K. acknowledges the hospitality of the Isaac Newton Institute, Cambridge, UK, during the workshop ``Complex Fluids in Evolving Domains'', and helpful discussions with Uwe Thiele. We would like to thank Debmalya Roy for assistance with the experiments.
\end{acknowledgments}

%
%
%


\begin{thebibliography}{69}%
\makeatletter
\providecommand \@ifxundefined [1]{%
 \@ifx{#1\undefined}
}%
\providecommand \@ifnum [1]{%
 \ifnum #1\expandafter \@firstoftwo
 \else \expandafter \@secondoftwo
 \fi
}%
\providecommand \@ifx [1]{%
 \ifx #1\expandafter \@firstoftwo
 \else \expandafter \@secondoftwo
 \fi
}%
\providecommand \natexlab [1]{#1}%
\providecommand \enquote  [1]{``#1''}%
\providecommand \bibnamefont  [1]{#1}%
\providecommand \bibfnamefont [1]{#1}%
\providecommand \citenamefont [1]{#1}%
\providecommand \href@noop [0]{\@secondoftwo}%
\providecommand \href [0]{\begingroup \@sanitize@url \@href}%
\providecommand \@href[1]{\@@startlink{#1}\@@href}%
\providecommand \@@href[1]{\endgroup#1\@@endlink}%
\providecommand \@sanitize@url [0]{\catcode `\\12\catcode `\$12\catcode
  `\&12\catcode `\#12\catcode `\^12\catcode `\_12\catcode `\%12\relax}%
\providecommand \@@startlink[1]{}%
\providecommand \@@endlink[0]{}%
\providecommand \url  [0]{\begingroup\@sanitize@url \@url }%
\providecommand \@url [1]{\endgroup\@href {#1}{\urlprefix }}%
\providecommand \urlprefix  [0]{URL }%
\providecommand \Eprint [0]{\href }%
\providecommand \doibase [0]{https://doi.org/}%
\providecommand \selectlanguage [0]{\@gobble}%
\providecommand \bibinfo  [0]{\@secondoftwo}%
\providecommand \bibfield  [0]{\@secondoftwo}%
\providecommand \translation [1]{[#1]}%
\providecommand \BibitemOpen [0]{}%
\providecommand \bibitemStop [0]{}%
\providecommand \bibitemNoStop [0]{.\EOS\space}%
\providecommand \EOS [0]{\spacefactor3000\relax}%
\providecommand \BibitemShut  [1]{\csname bibitem#1\endcsname}%
\let\auto@bib@innerbib\@empty
\bibitem [{\citenamefont {Lohse}\ and\ \citenamefont
  {Zhang}(2020)}]{Lohse:NRP2020}%
  \BibitemOpen
  \bibfield  {author} {\bibinfo {author} {\bibfnamefont {D.}~\bibnamefont
  {Lohse}}\ and\ \bibinfo {author} {\bibfnamefont {X.}~\bibnamefont {Zhang}},\
  }\bibfield  {title} {\bibinfo {title} {Physicochemical hydrodynamics of
  droplets out of equilibrium},\ }\href
  {https://doi.org/10.1038/s42254-020-0199-z} {\bibfield  {journal} {\bibinfo
  {journal} {Nature Reviews Physics}\ }\textbf {\bibinfo {volume} {2}},\
  \bibinfo {pages} {426} (\bibinfo {year} {2020})}\BibitemShut {NoStop}%
\bibitem [{\citenamefont {Brutin}\ and\ \citenamefont
  {Starov}(2018)}]{Brutin:ChemSocRev2018}%
  \BibitemOpen
  \bibfield  {author} {\bibinfo {author} {\bibfnamefont {D.}~\bibnamefont
  {Brutin}}\ and\ \bibinfo {author} {\bibfnamefont {V.}~\bibnamefont
  {Starov}},\ }\bibfield  {title} {\bibinfo {title} {Recent advances in droplet
  wetting and evaporation},\ }\href {https://doi.org/10.1039/C6CS00902F}
  {\bibfield  {journal} {\bibinfo  {journal} {Chemical Society Reviews}\
  }\textbf {\bibinfo {volume} {47}},\ \bibinfo {pages} {558} (\bibinfo {year}
  {2018})}\BibitemShut {NoStop}%
\bibitem [{\citenamefont {Snoeijer}\ and\ \citenamefont
  {Andreotti}(2013)}]{Snoeijer:ARFM2013}%
  \BibitemOpen
  \bibfield  {author} {\bibinfo {author} {\bibfnamefont {J.~H.}\ \bibnamefont
  {Snoeijer}}\ and\ \bibinfo {author} {\bibfnamefont {B.}~\bibnamefont
  {Andreotti}},\ }\bibfield  {title} {\bibinfo {title} {Moving contact lines:
  Scales, regimes, and dynamical transitions},\ }\href
  {https://doi.org/10.1146/annurev-fluid-011212-140734} {\bibfield  {journal}
  {\bibinfo  {journal} {Annual Review of Fluid Mechanics}\ }\textbf {\bibinfo
  {volume} {45}},\ \bibinfo {pages} {269} (\bibinfo {year} {2013})}\BibitemShut
  {NoStop}%
\bibitem [{\citenamefont {Bonn}\ \emph {et~al.}(2009)\citenamefont {Bonn},
  \citenamefont {Eggers}, \citenamefont {Indekeu}, \citenamefont {Meunier},\
  and\ \citenamefont {Rolley}}]{Bonn:RMP2009}%
  \BibitemOpen
  \bibfield  {author} {\bibinfo {author} {\bibfnamefont {D.}~\bibnamefont
  {Bonn}}, \bibinfo {author} {\bibfnamefont {J.}~\bibnamefont {Eggers}},
  \bibinfo {author} {\bibfnamefont {J.}~\bibnamefont {Indekeu}}, \bibinfo
  {author} {\bibfnamefont {J.}~\bibnamefont {Meunier}},\ and\ \bibinfo {author}
  {\bibfnamefont {E.}~\bibnamefont {Rolley}},\ }\bibfield  {title} {\bibinfo
  {title} {Wetting and spreading},\ }\href
  {https://doi.org/10.1103/revmodphys.81.739} {\bibfield  {journal} {\bibinfo
  {journal} {Rev. Mod. Phys.}\ }\textbf {\bibinfo {volume} {81}},\ \bibinfo
  {pages} {739} (\bibinfo {year} {2009})}\BibitemShut {NoStop}%
\bibitem [{\citenamefont {Smith}\ \emph {et~al.}(2018)\citenamefont {Smith},
  \citenamefont {Theodorakis}, \citenamefont {Craster},\ and\ \citenamefont
  {Matar}}]{Smith:L2018}%
  \BibitemOpen
  \bibfield  {author} {\bibinfo {author} {\bibfnamefont {E.~R.}\ \bibnamefont
  {Smith}}, \bibinfo {author} {\bibfnamefont {P.~E.}\ \bibnamefont
  {Theodorakis}}, \bibinfo {author} {\bibfnamefont {R.~V.}\ \bibnamefont
  {Craster}},\ and\ \bibinfo {author} {\bibfnamefont {O.~K.}\ \bibnamefont
  {Matar}},\ }\bibfield  {title} {\bibinfo {title} {Moving contact lines:
  Linking molecular dynamics and continuum-scale modeling},\ }\href
  {https://doi.org/10.1021/acs.langmuir.8b00466} {\bibfield  {journal}
  {\bibinfo  {journal} {Langmuir}\ }\textbf {\bibinfo {volume} {34}},\ \bibinfo
  {pages} {12501} (\bibinfo {year} {2018})}\BibitemShut {NoStop}%
\bibitem [{\citenamefont {Brutin}\ \emph {et~al.}(2010)\citenamefont {Brutin},
  \citenamefont {Sobac}, \citenamefont {Loquet},\ and\ \citenamefont
  {Sampol}}]{Brutin:JFM2010}%
  \BibitemOpen
  \bibfield  {author} {\bibinfo {author} {\bibfnamefont {D.}~\bibnamefont
  {Brutin}}, \bibinfo {author} {\bibfnamefont {B.}~\bibnamefont {Sobac}},
  \bibinfo {author} {\bibfnamefont {B.}~\bibnamefont {Loquet}},\ and\ \bibinfo
  {author} {\bibfnamefont {J.}~\bibnamefont {Sampol}},\ }\bibfield  {title}
  {\bibinfo {title} {Pattern formation in drying drops of blood},\ }\href
  {https://doi.org/10.1017/s0022112010005070} {\bibfield  {journal} {\bibinfo
  {journal} {Journal of Fluid Mechanics}\ }\textbf {\bibinfo {volume} {667}},\
  \bibinfo {pages} {85} (\bibinfo {year} {2010})}\BibitemShut {NoStop}%
\bibitem [{\citenamefont {Traipe-Castro}\ \emph {et~al.}(2014)\citenamefont
  {Traipe-Castro}, \citenamefont {Salinas-Toro}, \citenamefont {L\'opez},
  \citenamefont {Zanolli}, \citenamefont {Srur}, \citenamefont {Valenzuela},
  \citenamefont {C\'aceres}, \citenamefont {Toledo-Araya},\ and\ \citenamefont
  {L\'opez-Sol\'is}}]{Traipe-Castro:BR2014}%
  \BibitemOpen
  \bibfield  {author} {\bibinfo {author} {\bibfnamefont {L.}~\bibnamefont
  {Traipe-Castro}}, \bibinfo {author} {\bibfnamefont {D.}~\bibnamefont
  {Salinas-Toro}}, \bibinfo {author} {\bibfnamefont {D.}~\bibnamefont
  {L\'opez}}, \bibinfo {author} {\bibfnamefont {M.}~\bibnamefont {Zanolli}},
  \bibinfo {author} {\bibfnamefont {M.}~\bibnamefont {Srur}}, \bibinfo {author}
  {\bibfnamefont {F.}~\bibnamefont {Valenzuela}}, \bibinfo {author}
  {\bibfnamefont {A.}~\bibnamefont {C\'aceres}}, \bibinfo {author}
  {\bibfnamefont {H.}~\bibnamefont {Toledo-Araya}},\ and\ \bibinfo {author}
  {\bibfnamefont {R.}~\bibnamefont {L\'opez-Sol\'is}},\ }\bibfield  {title}
  {\bibinfo {title} {Dynamics of tear fluid desiccation on a glass surface: a
  contribution to tear quality assessment},\ }\bibfield  {journal} {\bibinfo
  {journal} {Biological Research}\ }\textbf {\bibinfo {volume} {47}},\ \href
  {https://doi.org/10.1186/0717-6287-47-25} {10.1186/0717-6287-47-25} (\bibinfo
  {year} {2014})\BibitemShut {NoStop}%
\bibitem [{\citenamefont {Gans}\ and\ \citenamefont
  {Schubert}(2004)}]{DeGans:L2004}%
  \BibitemOpen
  \bibfield  {author} {\bibinfo {author} {\bibfnamefont {B.~J.~D.}\
  \bibnamefont {Gans}}\ and\ \bibinfo {author} {\bibfnamefont {U.~S.}\
  \bibnamefont {Schubert}},\ }\bibfield  {title} {\bibinfo {title} {Inkjet
  printing of well-defined polymer dots and arrays},\ }\href
  {https://doi.org/10.1021/la049469o} {\bibfield  {journal} {\bibinfo
  {journal} {Langmuir}\ }\textbf {\bibinfo {volume} {20}},\ \bibinfo {pages}
  {7789} (\bibinfo {year} {2004})}\BibitemShut {NoStop}%
\bibitem [{\citenamefont {Park}\ and\ \citenamefont {Moon}(2006)}]{Park:L2006}%
  \BibitemOpen
  \bibfield  {author} {\bibinfo {author} {\bibfnamefont {J.}~\bibnamefont
  {Park}}\ and\ \bibinfo {author} {\bibfnamefont {J.}~\bibnamefont {Moon}},\
  }\bibfield  {title} {\bibinfo {title} {Control of colloidal particle deposit
  patterns within picoliter droplets ejected by ink-jet printing},\ }\href
  {https://doi.org/10.1021/la053450j} {\bibfield  {journal} {\bibinfo
  {journal} {Langmuir}\ }\textbf {\bibinfo {volume} {22}},\ \bibinfo {pages}
  {3506} (\bibinfo {year} {2006})}\BibitemShut {NoStop}%
\bibitem [{\citenamefont {Zenit}(2019)}]{Zenit:PRF2019}%
  \BibitemOpen
  \bibfield  {author} {\bibinfo {author} {\bibfnamefont {R.}~\bibnamefont
  {Zenit}},\ }\bibfield  {title} {\bibinfo {title} {Some fluid mechanical
  aspects of artistic painting},\ }\href
  {https://doi.org/10.1103/PhysRevFluids.4.110507} {\bibfield  {journal}
  {\bibinfo  {journal} {Physical Review Fluids}\ }\textbf {\bibinfo {volume}
  {4}},\ \bibinfo {pages} {110507} (\bibinfo {year} {2019})}\BibitemShut
  {NoStop}%
\bibitem [{\citenamefont {Giorgiutti-Dauphin\'e}\ and\ \citenamefont
  {Pauchard}(2016)}]{GiorgiuttiDauphine:JAP2016}%
  \BibitemOpen
  \bibfield  {author} {\bibinfo {author} {\bibfnamefont {F.}~\bibnamefont
  {Giorgiutti-Dauphin\'e}}\ and\ \bibinfo {author} {\bibfnamefont
  {L.}~\bibnamefont {Pauchard}},\ }\bibfield  {title} {\bibinfo {title}
  {Painting cracks: A way to investigate the pictorial matter},\ }\href
  {https://doi.org/10.1063/1.4960438} {\bibfield  {journal} {\bibinfo
  {journal} {Journal of Applied Physics}\ }\textbf {\bibinfo {volume} {120}},\
  \bibinfo {pages} {065107} (\bibinfo {year} {2016})}\BibitemShut {NoStop}%
\bibitem [{\citenamefont {Cira}\ \emph {et~al.}(2015)\citenamefont {Cira},
  \citenamefont {Benusiglio},\ and\ \citenamefont {Prakash}}]{Cira:N2015}%
  \BibitemOpen
  \bibfield  {author} {\bibinfo {author} {\bibfnamefont {N.~J.}\ \bibnamefont
  {Cira}}, \bibinfo {author} {\bibfnamefont {A.}~\bibnamefont {Benusiglio}},\
  and\ \bibinfo {author} {\bibfnamefont {M.}~\bibnamefont {Prakash}},\
  }\bibfield  {title} {\bibinfo {title} {Vapour-mediated sensing and motility
  in two-component droplets},\ }\href {https://doi.org/10.1038/nature14272}
  {\bibfield  {journal} {\bibinfo  {journal} {Nature}\ }\textbf {\bibinfo
  {volume} {519}},\ \bibinfo {pages} {446} (\bibinfo {year}
  {2015})}\BibitemShut {NoStop}%
\bibitem [{\citenamefont {Karpitschka}\ \emph {et~al.}(2017)\citenamefont
  {Karpitschka}, \citenamefont {Liebig},\ and\ \citenamefont
  {Riegler}}]{Karpitschka:L2017}%
  \BibitemOpen
  \bibfield  {author} {\bibinfo {author} {\bibfnamefont {S.}~\bibnamefont
  {Karpitschka}}, \bibinfo {author} {\bibfnamefont {F.}~\bibnamefont
  {Liebig}},\ and\ \bibinfo {author} {\bibfnamefont {H.}~\bibnamefont
  {Riegler}},\ }\bibfield  {title} {\bibinfo {title} {Marangoni contraction of
  evaporating sessile droplets of binary mixtures},\ }\href
  {https://doi.org/10.1021/acs.langmuir.7b00740} {\bibfield  {journal}
  {\bibinfo  {journal} {Langmuir}\ }\textbf {\bibinfo {volume} {33}},\ \bibinfo
  {pages} {4682} (\bibinfo {year} {2017})}\BibitemShut {NoStop}%
\bibitem [{\citenamefont {Benusiglio}\ \emph {et~al.}(2018)\citenamefont
  {Benusiglio}, \citenamefont {Cira},\ and\ \citenamefont
  {Prakash}}]{Benusiglio:SM2018}%
  \BibitemOpen
  \bibfield  {author} {\bibinfo {author} {\bibfnamefont {A.}~\bibnamefont
  {Benusiglio}}, \bibinfo {author} {\bibfnamefont {N.~J.}\ \bibnamefont
  {Cira}},\ and\ \bibinfo {author} {\bibfnamefont {M.}~\bibnamefont
  {Prakash}},\ }\bibfield  {title} {\bibinfo {title} {Two-component
  marangoni-contracted droplets: friction and shape},\ }\href
  {https://doi.org/10.1039/c7sm02361h} {\bibfield  {journal} {\bibinfo
  {journal} {Soft Matter}\ }\textbf {\bibinfo {volume} {14}},\ \bibinfo {pages}
  {7724} (\bibinfo {year} {2018})}\BibitemShut {NoStop}%
\bibitem [{\citenamefont {Sadafi}\ \emph {et~al.}(2019)\citenamefont {Sadafi},
  \citenamefont {Dehaeck}, \citenamefont {Rednikov},\ and\ \citenamefont
  {Colinet}}]{Sadafi:L2019}%
  \BibitemOpen
  \bibfield  {author} {\bibinfo {author} {\bibfnamefont {H.}~\bibnamefont
  {Sadafi}}, \bibinfo {author} {\bibfnamefont {S.}~\bibnamefont {Dehaeck}},
  \bibinfo {author} {\bibfnamefont {A.}~\bibnamefont {Rednikov}},\ and\
  \bibinfo {author} {\bibfnamefont {P.}~\bibnamefont {Colinet}},\ }\bibfield
  {title} {\bibinfo {title} {Vapor-mediated versus substrate-mediated
  interactions between volatile droplets},\ }\href
  {https://doi.org/10.1021/acs.langmuir.9b00522} {\bibfield  {journal}
  {\bibinfo  {journal} {Langmuir}\ }\textbf {\bibinfo {volume} {35}},\ \bibinfo
  {pages} {7060} (\bibinfo {year} {2019})}\BibitemShut {NoStop}%
\bibitem [{\citenamefont {Malinowski}\ \emph {et~al.}(2020)\citenamefont
  {Malinowski}, \citenamefont {Parkin},\ and\ \citenamefont
  {Volpe}}]{Malinowski:SA2020}%
  \BibitemOpen
  \bibfield  {author} {\bibinfo {author} {\bibfnamefont {R.}~\bibnamefont
  {Malinowski}}, \bibinfo {author} {\bibfnamefont {I.~P.}\ \bibnamefont
  {Parkin}},\ and\ \bibinfo {author} {\bibfnamefont {G.}~\bibnamefont
  {Volpe}},\ }\bibfield  {title} {\bibinfo {title} {Nonmonotonic contactless
  manipulation of binary droplets via sensing of localized vapor sources on
  pristine substrates},\ }\bibfield  {journal} {\bibinfo  {journal} {Science
  Advances}\ }\textbf {\bibinfo {volume} {6}},\ \href
  {https://doi.org/10.1126/sciadv.aba3636} {10.1126/sciadv.aba3636} (\bibinfo
  {year} {2020})\BibitemShut {NoStop}%
\bibitem [{\citenamefont {Williams}\ \emph {et~al.}(2020)\citenamefont
  {Williams}, \citenamefont {Karapetsas}, \citenamefont {Mamalis},
  \citenamefont {Sefiane}, \citenamefont {Matar},\ and\ \citenamefont
  {Valluri}}]{Williams:JFM2020}%
  \BibitemOpen
  \bibfield  {author} {\bibinfo {author} {\bibfnamefont {A.~G.~L.}\
  \bibnamefont {Williams}}, \bibinfo {author} {\bibfnamefont {G.}~\bibnamefont
  {Karapetsas}}, \bibinfo {author} {\bibfnamefont {D.}~\bibnamefont {Mamalis}},
  \bibinfo {author} {\bibfnamefont {K.}~\bibnamefont {Sefiane}}, \bibinfo
  {author} {\bibfnamefont {O.~K.}\ \bibnamefont {Matar}},\ and\ \bibinfo
  {author} {\bibfnamefont {P.}~\bibnamefont {Valluri}},\ }\bibfield  {title}
  {\bibinfo {title} {Spreading and retraction dynamics of sessile evaporating
  droplets comprising volatile binary mixtures},\ }\bibfield  {journal}
  {\bibinfo  {journal} {Journal of Fluid Mechanics}\ }\textbf {\bibinfo
  {volume} {907}},\ \href {https://doi.org/10.1017/jfm.2020.840}
  {10.1017/jfm.2020.840} (\bibinfo {year} {2020})\BibitemShut {NoStop}%
\bibitem [{\citenamefont {Hack}\ \emph {et~al.}(2021)\citenamefont {Hack},
  \citenamefont {Kwieciński}, \citenamefont {Ramírez-Soto}, \citenamefont
  {Segers}, \citenamefont {Karpitschka}, \citenamefont {Kooij},\ and\
  \citenamefont {Snoeijer}}]{Hack:L2021}%
  \BibitemOpen
  \bibfield  {author} {\bibinfo {author} {\bibfnamefont {M.~A.}\ \bibnamefont
  {Hack}}, \bibinfo {author} {\bibfnamefont {W.}~\bibnamefont {Kwieciński}},
  \bibinfo {author} {\bibfnamefont {O.}~\bibnamefont {Ramírez-Soto}}, \bibinfo
  {author} {\bibfnamefont {T.}~\bibnamefont {Segers}}, \bibinfo {author}
  {\bibfnamefont {S.}~\bibnamefont {Karpitschka}}, \bibinfo {author}
  {\bibfnamefont {E.~S.}\ \bibnamefont {Kooij}},\ and\ \bibinfo {author}
  {\bibfnamefont {J.~H.}\ \bibnamefont {Snoeijer}},\ }\bibfield  {title}
  {\bibinfo {title} {Wetting of two-component drops: Marangoni contraction
  versus autophobing},\ }\href {https://doi.org/10.1021/acs.langmuir.0c03571}
  {\bibfield  {journal} {\bibinfo  {journal} {Langmuir}\ }\textbf {\bibinfo
  {volume} {37}},\ \bibinfo {pages} {3605} (\bibinfo {year} {2021})},\ \bibinfo
  {note} {pMID: 33734702},\ \Eprint
  {https://arxiv.org/abs/https://doi.org/10.1021/acs.langmuir.0c03571}
  {https://doi.org/10.1021/acs.langmuir.0c03571} \BibitemShut {NoStop}%
\bibitem [{\citenamefont {Shiri}\ \emph {et~al.}(2021)\citenamefont {Shiri},
  \citenamefont {Sinha}, \citenamefont {Baumgartner},\ and\ \citenamefont
  {Cira}}]{Shiri:PRL2021}%
  \BibitemOpen
  \bibfield  {author} {\bibinfo {author} {\bibfnamefont {S.}~\bibnamefont
  {Shiri}}, \bibinfo {author} {\bibfnamefont {S.}~\bibnamefont {Sinha}},
  \bibinfo {author} {\bibfnamefont {D.~A.}\ \bibnamefont {Baumgartner}},\ and\
  \bibinfo {author} {\bibfnamefont {N.~J.}\ \bibnamefont {Cira}},\ }\bibfield
  {title} {\bibinfo {title} {Thermal marangoni flow impacts the shape of single
  component volatile droplets on thin, completely wetting substrates},\ }\href
  {https://doi.org/10.1103/physrevlett.127.024502} {\bibfield  {journal}
  {\bibinfo  {journal} {Physical Review Letters}\ }\textbf {\bibinfo {volume}
  {127}},\ \bibinfo {pages} {024502} (\bibinfo {year} {2021})}\BibitemShut
  {NoStop}%
\bibitem [{\citenamefont {Darhuber}\ and\ \citenamefont
  {Troian}(2003)}]{Darhuber:PoF2003}%
  \BibitemOpen
  \bibfield  {author} {\bibinfo {author} {\bibfnamefont {A.~A.}\ \bibnamefont
  {Darhuber}}\ and\ \bibinfo {author} {\bibfnamefont {S.~M.}\ \bibnamefont
  {Troian}},\ }\bibfield  {title} {\bibinfo {title} {Marangoni driven
  structures in thin film flows},\ }\href {https://doi.org/10.1063/1.4739213}
  {\bibfield  {journal} {\bibinfo  {journal} {Physics of Fluids}\ }\textbf
  {\bibinfo {volume} {15}},\ \bibinfo {pages} {S9} (\bibinfo {year} {2003})},\
  \Eprint {https://arxiv.org/abs/https://doi.org/10.1063/1.4739213}
  {https://doi.org/10.1063/1.4739213} \BibitemShut {NoStop}%
\bibitem [{\citenamefont {Gotkis}\ \emph {et~al.}(2006)\citenamefont {Gotkis},
  \citenamefont {Ivanov}, \citenamefont {Murisic},\ and\ \citenamefont
  {Kondic}}]{Gotkis:PRL2006}%
  \BibitemOpen
  \bibfield  {author} {\bibinfo {author} {\bibfnamefont {Y.}~\bibnamefont
  {Gotkis}}, \bibinfo {author} {\bibfnamefont {I.}~\bibnamefont {Ivanov}},
  \bibinfo {author} {\bibfnamefont {N.}~\bibnamefont {Murisic}},\ and\ \bibinfo
  {author} {\bibfnamefont {L.}~\bibnamefont {Kondic}},\ }\bibfield  {title}
  {\bibinfo {title} {Dynamic structure formation at the fronts of volatile
  liquid drops},\ }\href {https://doi.org/10.1103/PhysRevLett.97.186101}
  {\bibfield  {journal} {\bibinfo  {journal} {Phys. Rev. Lett.}\ }\textbf
  {\bibinfo {volume} {97}},\ \bibinfo {pages} {186101} (\bibinfo {year}
  {2006})}\BibitemShut {NoStop}%
\bibitem [{\citenamefont {Wodlei}\ \emph {et~al.}(2018)\citenamefont {Wodlei},
  \citenamefont {Sebilleau}, \citenamefont {Magnaudet},\ and\ \citenamefont
  {Pimienta}}]{Wodlei:NC2018}%
  \BibitemOpen
  \bibfield  {author} {\bibinfo {author} {\bibfnamefont {F.}~\bibnamefont
  {Wodlei}}, \bibinfo {author} {\bibfnamefont {J.}~\bibnamefont {Sebilleau}},
  \bibinfo {author} {\bibfnamefont {J.}~\bibnamefont {Magnaudet}},\ and\
  \bibinfo {author} {\bibfnamefont {V.}~\bibnamefont {Pimienta}},\ }\bibfield
  {title} {\bibinfo {title} {Marangoni-driven flower-like patterning of an
  evaporating drop spreading on a liquid substrate},\ }\bibfield  {journal}
  {\bibinfo  {journal} {Nature Communications}\ }\textbf {\bibinfo {volume}
  {9}},\ \href {https://doi.org/10.1038/s41467-018-03201-3}
  {10.1038/s41467-018-03201-3} (\bibinfo {year} {2018})\BibitemShut {NoStop}%
\bibitem [{\citenamefont {Oron}\ \emph {et~al.}(1997)\citenamefont {Oron},
  \citenamefont {Davis},\ and\ \citenamefont {Bankoff}}]{Oron:RMP1997}%
  \BibitemOpen
  \bibfield  {author} {\bibinfo {author} {\bibfnamefont {A.}~\bibnamefont
  {Oron}}, \bibinfo {author} {\bibfnamefont {S.~H.}\ \bibnamefont {Davis}},\
  and\ \bibinfo {author} {\bibfnamefont {S.~G.}\ \bibnamefont {Bankoff}},\
  }\bibfield  {title} {\bibinfo {title} {Long-scale evolution of thin liquid
  films},\ }\href {https://doi.org/10.1103/revmodphys.69.931} {\bibfield
  {journal} {\bibinfo  {journal} {Reviews of Modern Physics}\ }\textbf
  {\bibinfo {volume} {69}},\ \bibinfo {pages} {931} (\bibinfo {year}
  {1997})}\BibitemShut {NoStop}%
\bibitem [{\citenamefont {Zhang}\ \emph {et~al.}(2011)\citenamefont {Zhang},
  \citenamefont {Oron},\ and\ \citenamefont {Behringer}}]{Zhang:PoF2011}%
  \BibitemOpen
  \bibfield  {author} {\bibinfo {author} {\bibfnamefont {J.}~\bibnamefont
  {Zhang}}, \bibinfo {author} {\bibfnamefont {A.}~\bibnamefont {Oron}},\ and\
  \bibinfo {author} {\bibfnamefont {R.~P.}\ \bibnamefont {Behringer}},\
  }\bibfield  {title} {\bibinfo {title} {Novel pattern forming states for
  marangoni convection in volatile binary liquids},\ }\href
  {https://doi.org/10.1063/1.3609287} {\bibfield  {journal} {\bibinfo
  {journal} {Physics of Fluids}\ }\textbf {\bibinfo {volume} {23}},\ \bibinfo
  {pages} {072102} (\bibinfo {year} {2011})}\BibitemShut {NoStop}%
\bibitem [{\citenamefont {Christy}\ \emph {et~al.}(2011)\citenamefont
  {Christy}, \citenamefont {Hamamoto},\ and\ \citenamefont
  {Sefiane}}]{Christy:PRL2011}%
  \BibitemOpen
  \bibfield  {author} {\bibinfo {author} {\bibfnamefont {J.~R.~E.}\
  \bibnamefont {Christy}}, \bibinfo {author} {\bibfnamefont {Y.}~\bibnamefont
  {Hamamoto}},\ and\ \bibinfo {author} {\bibfnamefont {K.}~\bibnamefont
  {Sefiane}},\ }\bibfield  {title} {\bibinfo {title} {Flow transition within an
  evaporating binary mixture sessile drop},\ }\href
  {https://doi.org/10.1103/physrevlett.106.205701} {\bibfield  {journal}
  {\bibinfo  {journal} {Physical Review Letters}\ }\textbf {\bibinfo {volume}
  {106}},\ \bibinfo {pages} {205701} (\bibinfo {year} {2011})}\BibitemShut
  {NoStop}%
\bibitem [{\citenamefont {Soulie}\ \emph {et~al.}(2015)\citenamefont {Soulie},
  \citenamefont {Karpitschka}, \citenamefont {Lequien}, \citenamefont
  {Pren\'e}, \citenamefont {Zemb}, \citenamefont {M\"ohwald},\ and\
  \citenamefont {Riegler}}]{Soulie:PCCP2015}%
  \BibitemOpen
  \bibfield  {author} {\bibinfo {author} {\bibfnamefont {V.}~\bibnamefont
  {Soulie}}, \bibinfo {author} {\bibfnamefont {S.}~\bibnamefont {Karpitschka}},
  \bibinfo {author} {\bibfnamefont {F.}~\bibnamefont {Lequien}}, \bibinfo
  {author} {\bibfnamefont {P.}~\bibnamefont {Pren\'e}}, \bibinfo {author}
  {\bibfnamefont {T.}~\bibnamefont {Zemb}}, \bibinfo {author} {\bibfnamefont
  {H.}~\bibnamefont {M\"ohwald}},\ and\ \bibinfo {author} {\bibfnamefont
  {H.}~\bibnamefont {Riegler}},\ }\bibfield  {title} {\bibinfo {title} {The
  evaporation behavior of sessile droplets from aqueous saline solutions},\
  }\href {https://doi.org/10.1039/C5CP02444G} {\bibfield  {journal} {\bibinfo
  {journal} {Phys. Chem. Chem. Phys.}\ }\textbf {\bibinfo {volume} {17}},\
  \bibinfo {pages} {22296} (\bibinfo {year} {2015})}\BibitemShut {NoStop}%
\bibitem [{\citenamefont {Marin}\ \emph {et~al.}(2019)\citenamefont {Marin},
  \citenamefont {Karpitschka}, \citenamefont {Noguera-Mar{\'{\i}}n},
  \citenamefont {Cabrerizo-V{\'{\i}}lchez}, \citenamefont {Rossi},
  \citenamefont {K\"ahler},\ and\ \citenamefont {{Rodr{\'{\i}}guez
  Valverde}}}]{Marin:PRF2019}%
  \BibitemOpen
  \bibfield  {author} {\bibinfo {author} {\bibfnamefont {A.}~\bibnamefont
  {Marin}}, \bibinfo {author} {\bibfnamefont {S.}~\bibnamefont {Karpitschka}},
  \bibinfo {author} {\bibfnamefont {D.}~\bibnamefont {Noguera-Mar{\'{\i}}n}},
  \bibinfo {author} {\bibfnamefont {M.~A.}\ \bibnamefont
  {Cabrerizo-V{\'{\i}}lchez}}, \bibinfo {author} {\bibfnamefont
  {M.}~\bibnamefont {Rossi}}, \bibinfo {author} {\bibfnamefont {C.~J.}\
  \bibnamefont {K\"ahler}},\ and\ \bibinfo {author} {\bibfnamefont {M.~A.}\
  \bibnamefont {{Rodr{\'{\i}}guez Valverde}}},\ }\bibfield  {title} {\bibinfo
  {title} {Solutal marangoni flow as the cause of ring stains from drying salty
  colloidal drops},\ }\href {https://doi.org/10.1103/physrevfluids.4.041601}
  {\bibfield  {journal} {\bibinfo  {journal} {Phys. Rev. Fluids}\ }\textbf
  {\bibinfo {volume} {4}},\ \bibinfo {pages} {041601(R)} (\bibinfo {year}
  {2019})}\BibitemShut {NoStop}%
\bibitem [{\citenamefont {Rossi}\ \emph {et~al.}(2019)\citenamefont {Rossi},
  \citenamefont {Marin},\ and\ \citenamefont {K\"ahler}}]{Rossi:PRE2019}%
  \BibitemOpen
  \bibfield  {author} {\bibinfo {author} {\bibfnamefont {M.}~\bibnamefont
  {Rossi}}, \bibinfo {author} {\bibfnamefont {A.}~\bibnamefont {Marin}},\ and\
  \bibinfo {author} {\bibfnamefont {C.~J.}\ \bibnamefont {K\"ahler}},\
  }\bibfield  {title} {\bibinfo {title} {Interfacial flows in sessile
  evaporating droplets of mineral water},\ }\href
  {https://doi.org/10.1103/physreve.100.033103} {\bibfield  {journal} {\bibinfo
   {journal} {Physical Review E}\ }\textbf {\bibinfo {volume} {100}},\ \bibinfo
  {pages} {033103} (\bibinfo {year} {2019})}\BibitemShut {NoStop}%
\bibitem [{\citenamefont {Karapetsas}\ \emph {et~al.}(2016)\citenamefont
  {Karapetsas}, \citenamefont {Sahu},\ and\ \citenamefont
  {Matar}}]{Karapetsas:L2016}%
  \BibitemOpen
  \bibfield  {author} {\bibinfo {author} {\bibfnamefont {G.}~\bibnamefont
  {Karapetsas}}, \bibinfo {author} {\bibfnamefont {K.~C.}\ \bibnamefont
  {Sahu}},\ and\ \bibinfo {author} {\bibfnamefont {O.~K.}\ \bibnamefont
  {Matar}},\ }\bibfield  {title} {\bibinfo {title} {Evaporation of sessile
  droplets laden with particles and insoluble surfactants},\ }\href
  {https://doi.org/10.1021/acs.langmuir.6b01042} {\bibfield  {journal}
  {\bibinfo  {journal} {Langmuir}\ }\textbf {\bibinfo {volume} {32}},\ \bibinfo
  {pages} {6871} (\bibinfo {year} {2016})}\BibitemShut {NoStop}%
\bibitem [{\citenamefont {Diddens}(2017)}]{Diddens:JCP2017}%
  \BibitemOpen
  \bibfield  {author} {\bibinfo {author} {\bibfnamefont {C.}~\bibnamefont
  {Diddens}},\ }\bibfield  {title} {\bibinfo {title} {Detailed finite element
  method modeling of evaporating multi-component droplets},\ }\href
  {https://doi.org/10.1016/j.jcp.2017.03.049} {\bibfield  {journal} {\bibinfo
  {journal} {Journal of Computational Physics}\ }\textbf {\bibinfo {volume}
  {340}},\ \bibinfo {pages} {670} (\bibinfo {year} {2017})}\BibitemShut
  {NoStop}%
\bibitem [{\citenamefont {Diddens}\ \emph {et~al.}(2017)\citenamefont
  {Diddens}, \citenamefont {Tan}, \citenamefont {Lv}, \citenamefont {Versluis},
  \citenamefont {Kuerten}, \citenamefont {Zhang},\ and\ \citenamefont
  {Lohse}}]{Diddens:JFM2017}%
  \BibitemOpen
  \bibfield  {author} {\bibinfo {author} {\bibfnamefont {C.}~\bibnamefont
  {Diddens}}, \bibinfo {author} {\bibfnamefont {H.}~\bibnamefont {Tan}},
  \bibinfo {author} {\bibfnamefont {P.}~\bibnamefont {Lv}}, \bibinfo {author}
  {\bibfnamefont {M.}~\bibnamefont {Versluis}}, \bibinfo {author}
  {\bibfnamefont {J.~G.~M.}\ \bibnamefont {Kuerten}}, \bibinfo {author}
  {\bibfnamefont {X.}~\bibnamefont {Zhang}},\ and\ \bibinfo {author}
  {\bibfnamefont {D.}~\bibnamefont {Lohse}},\ }\bibfield  {title} {\bibinfo
  {title} {Evaporating pure, binary and ternary droplets: thermal effects and
  axial symmetry breaking},\ }\href {https://doi.org/10.1017/jfm.2017.312}
  {\bibfield  {journal} {\bibinfo  {journal} {Journal of Fluid Mechanics}\
  }\textbf {\bibinfo {volume} {823}},\ \bibinfo {pages} {470} (\bibinfo {year}
  {2017})}\BibitemShut {NoStop}%
\bibitem [{\citenamefont {Li}\ \emph {et~al.}(2019)\citenamefont {Li},
  \citenamefont {Diddens}, \citenamefont {Lv}, \citenamefont {Wijshoff},
  \citenamefont {Versluis},\ and\ \citenamefont {Lohse}}]{Li:PRL2019}%
  \BibitemOpen
  \bibfield  {author} {\bibinfo {author} {\bibfnamefont {Y.}~\bibnamefont
  {Li}}, \bibinfo {author} {\bibfnamefont {C.}~\bibnamefont {Diddens}},
  \bibinfo {author} {\bibfnamefont {P.}~\bibnamefont {Lv}}, \bibinfo {author}
  {\bibfnamefont {H.}~\bibnamefont {Wijshoff}}, \bibinfo {author}
  {\bibfnamefont {M.}~\bibnamefont {Versluis}},\ and\ \bibinfo {author}
  {\bibfnamefont {D.}~\bibnamefont {Lohse}},\ }\bibfield  {title} {\bibinfo
  {title} {Gravitational effect in evaporating binary microdroplets},\ }\href
  {https://doi.org/10.1103/physrevlett.122.114501} {\bibfield  {journal}
  {\bibinfo  {journal} {Physical Review Letters}\ }\textbf {\bibinfo {volume}
  {122}},\ \bibinfo {pages} {114501} (\bibinfo {year} {2019})}\BibitemShut
  {NoStop}%
\bibitem [{\citenamefont {van Gaalen}\ \emph {et~al.}(2021)\citenamefont {van
  Gaalen}, \citenamefont {Diddens}, \citenamefont {Wijshoff},\ and\
  \citenamefont {Kuerten}}]{Gaalen:JCIS2021}%
  \BibitemOpen
  \bibfield  {author} {\bibinfo {author} {\bibfnamefont {R.~T.}\ \bibnamefont
  {van Gaalen}}, \bibinfo {author} {\bibfnamefont {C.}~\bibnamefont {Diddens}},
  \bibinfo {author} {\bibfnamefont {H.~M.~A.}\ \bibnamefont {Wijshoff}},\ and\
  \bibinfo {author} {\bibfnamefont {J.~G.~M.}\ \bibnamefont {Kuerten}},\
  }\bibfield  {title} {\bibinfo {title} {Marangoni circulation in evaporating
  droplets in the presence of soluble surfactants},\ }\href
  {https://doi.org/10.1016/j.jcis.2020.10.057} {\bibfield  {journal} {\bibinfo
  {journal} {Journal of Colloid and Interface Science}\ }\textbf {\bibinfo
  {volume} {584}},\ \bibinfo {pages} {622} (\bibinfo {year}
  {2021})}\BibitemShut {NoStop}%
\bibitem [{\citenamefont {Pahlavan}\ \emph {et~al.}(2021)\citenamefont
  {Pahlavan}, \citenamefont {Yang}, \citenamefont {Bain},\ and\ \citenamefont
  {Stone}}]{Pahlavan:PRL2021}%
  \BibitemOpen
  \bibfield  {author} {\bibinfo {author} {\bibfnamefont {A.~A.}\ \bibnamefont
  {Pahlavan}}, \bibinfo {author} {\bibfnamefont {L.}~\bibnamefont {Yang}},
  \bibinfo {author} {\bibfnamefont {C.~D.}\ \bibnamefont {Bain}},\ and\
  \bibinfo {author} {\bibfnamefont {H.~A.}\ \bibnamefont {Stone}},\ }\bibfield
  {title} {\bibinfo {title} {Evaporation of binary-mixture liquid droplets: The
  formation of picoliter pancakelike shapes},\ }\href
  {https://doi.org/10.1103/physrevlett.127.024501} {\bibfield  {journal}
  {\bibinfo  {journal} {Physical Review Letters}\ }\textbf {\bibinfo {volume}
  {127}},\ \bibinfo {pages} {024501} (\bibinfo {year} {2021})}\BibitemShut
  {NoStop}%
\bibitem [{\citenamefont {Craster}\ and\ \citenamefont
  {Matar}(2009)}]{Craster:RMP2009}%
  \BibitemOpen
  \bibfield  {author} {\bibinfo {author} {\bibfnamefont {R.~V.}\ \bibnamefont
  {Craster}}\ and\ \bibinfo {author} {\bibfnamefont {O.~K.}\ \bibnamefont
  {Matar}},\ }\bibfield  {title} {\bibinfo {title} {Dynamics and stability of
  thin liquid films},\ }\href {https://doi.org/10.1103/revmodphys.81.1131}
  {\bibfield  {journal} {\bibinfo  {journal} {Reviews of Modern Physics}\
  }\textbf {\bibinfo {volume} {81}},\ \bibinfo {pages} {1131} (\bibinfo {year}
  {2009})}\BibitemShut {NoStop}%
\bibitem [{\citenamefont {Thiele}\ \emph {et~al.}(2012)\citenamefont {Thiele},
  \citenamefont {Archer},\ and\ \citenamefont {Plapp}}]{Thiele:PoF2012}%
  \BibitemOpen
  \bibfield  {author} {\bibinfo {author} {\bibfnamefont {U.}~\bibnamefont
  {Thiele}}, \bibinfo {author} {\bibfnamefont {A.~J.}\ \bibnamefont {Archer}},\
  and\ \bibinfo {author} {\bibfnamefont {M.}~\bibnamefont {Plapp}},\ }\bibfield
   {title} {\bibinfo {title} {Thermodynamically consistent description of the
  hydrodynamics of free surfaces covered by insoluble surfactants of high
  concentration},\ }\href {https://doi.org/10.1063/1.4758476} {\bibfield
  {journal} {\bibinfo  {journal} {Physics of Fluids}\ }\textbf {\bibinfo
  {volume} {24}},\ \bibinfo {pages} {102107} (\bibinfo {year}
  {2012})}\BibitemShut {NoStop}%
\bibitem [{\citenamefont {Thiele}\ \emph {et~al.}(2016)\citenamefont {Thiele},
  \citenamefont {Archer},\ and\ \citenamefont {Pismen}}]{Thiele:PRF2016}%
  \BibitemOpen
  \bibfield  {author} {\bibinfo {author} {\bibfnamefont {U.}~\bibnamefont
  {Thiele}}, \bibinfo {author} {\bibfnamefont {A.~J.}\ \bibnamefont {Archer}},\
  and\ \bibinfo {author} {\bibfnamefont {L.~M.}\ \bibnamefont {Pismen}},\
  }\bibfield  {title} {\bibinfo {title} {Gradient dynamics models for liquid
  films with soluble surfactant},\ }\href
  {https://doi.org/10.1103/physrevfluids.1.083903} {\bibfield  {journal}
  {\bibinfo  {journal} {Physical Review Fluids}\ }\textbf {\bibinfo {volume}
  {1}},\ \bibinfo {pages} {083903} (\bibinfo {year} {2016})}\BibitemShut
  {NoStop}%
\bibitem [{\citenamefont {Xu}\ \emph {et~al.}(2015)\citenamefont {Xu},
  \citenamefont {Thiele},\ and\ \citenamefont {Qian}}]{Xu:JPCM2015}%
  \BibitemOpen
  \bibfield  {author} {\bibinfo {author} {\bibfnamefont {X.}~\bibnamefont
  {Xu}}, \bibinfo {author} {\bibfnamefont {U.}~\bibnamefont {Thiele}},\ and\
  \bibinfo {author} {\bibfnamefont {T.}~\bibnamefont {Qian}},\ }\bibfield
  {title} {\bibinfo {title} {A variational approach to thin film hydrodynamics
  of binary mixtures},\ }\href {https://doi.org/10.1088/0953-8984/27/8/085005}
  {\bibfield  {journal} {\bibinfo  {journal} {Journal of Physics: Condensed
  Matter}\ }\textbf {\bibinfo {volume} {27}},\ \bibinfo {pages} {085005}
  (\bibinfo {year} {2015})}\BibitemShut {NoStop}%
\bibitem [{\citenamefont {Moshinskii}(2004)}]{Moshinskii:FD2004}%
  \BibitemOpen
  \bibfield  {author} {\bibinfo {author} {\bibfnamefont {A.~I.}\ \bibnamefont
  {Moshinskii}},\ }\bibfield  {title} {\bibinfo {title} {Equivalent diffusion
  in a thin film flow with a non-one-dimensional velocity field and anisotropic
  diffusion tensor},\ }\href
  {https://doi.org/10.1023/b:flui.0000030307.90317.f6} {\bibfield  {journal}
  {\bibinfo  {journal} {Fluid Dynamics}\ }\textbf {\bibinfo {volume} {39}},\
  \bibinfo {pages} {230} (\bibinfo {year} {2004})}\BibitemShut {NoStop}%
\bibitem [{\citenamefont {Ajdari}\ \emph {et~al.}(2006)\citenamefont {Ajdari},
  \citenamefont {Bontoux},\ and\ \citenamefont {Stone}}]{Ajdari:AC2006}%
  \BibitemOpen
  \bibfield  {author} {\bibinfo {author} {\bibfnamefont {A.}~\bibnamefont
  {Ajdari}}, \bibinfo {author} {\bibfnamefont {N.}~\bibnamefont {Bontoux}},\
  and\ \bibinfo {author} {\bibfnamefont {H.~A.}\ \bibnamefont {Stone}},\
  }\bibfield  {title} {\bibinfo {title} {Hydrodynamic dispersion in shallow
  microchannels:~ the effect of cross-sectional shape},\ }\href
  {https://doi.org/10.1021/ac0508651} {\bibfield  {journal} {\bibinfo
  {journal} {Analytical Chemistry}\ }\textbf {\bibinfo {volume} {78}},\
  \bibinfo {pages} {387} (\bibinfo {year} {2006})}\BibitemShut {NoStop}%
\bibitem [{\citenamefont {Mukahal}\ \emph {et~al.}(2017)\citenamefont
  {Mukahal}, \citenamefont {Duffy},\ and\ \citenamefont
  {Wilson}}]{Mukahal:PRSA2017}%
  \BibitemOpen
  \bibfield  {author} {\bibinfo {author} {\bibfnamefont {F.~H. H.~A.}\
  \bibnamefont {Mukahal}}, \bibinfo {author} {\bibfnamefont {B.~R.}\
  \bibnamefont {Duffy}},\ and\ \bibinfo {author} {\bibfnamefont {S.~K.}\
  \bibnamefont {Wilson}},\ }\bibfield  {title} {\bibinfo {title} {Advection and
  taylor{\textendash}aris dispersion in rivulet flow},\ }\href
  {https://doi.org/10.1098/rspa.2017.0524} {\bibfield  {journal} {\bibinfo
  {journal} {Proceedings of the Royal Society A: Mathematical, Physical and
  Engineering Sciences}\ }\textbf {\bibinfo {volume} {473}},\ \bibinfo {pages}
  {20170524} (\bibinfo {year} {2017})}\BibitemShut {NoStop}%
\bibitem [{\citenamefont {Vilquin}\ \emph {et~al.}(2020)\citenamefont
  {Vilquin}, \citenamefont {Bertin}, \citenamefont {Soulard}, \citenamefont
  {Guyard}, \citenamefont {Raphael}, \citenamefont {Restagno}, \citenamefont
  {Salez},\ and\ \citenamefont {McGraw}}]{Vilquin:a2020}%
  \BibitemOpen
  \bibfield  {author} {\bibinfo {author} {\bibfnamefont {A.}~\bibnamefont
  {Vilquin}}, \bibinfo {author} {\bibfnamefont {V.}~\bibnamefont {Bertin}},
  \bibinfo {author} {\bibfnamefont {P.}~\bibnamefont {Soulard}}, \bibinfo
  {author} {\bibfnamefont {G.}~\bibnamefont {Guyard}}, \bibinfo {author}
  {\bibfnamefont {E.}~\bibnamefont {Raphael}}, \bibinfo {author} {\bibfnamefont
  {F.}~\bibnamefont {Restagno}}, \bibinfo {author} {\bibfnamefont
  {T.}~\bibnamefont {Salez}},\ and\ \bibinfo {author} {\bibfnamefont {J.~D.}\
  \bibnamefont {McGraw}},\ }\bibfield  {title} {\bibinfo {title} {Time
  dependence of advection-diffusion coupling for nanoparticle ensembles},\
  }\href@noop {} {\bibfield  {journal} {\bibinfo  {journal} {Phys. Rev. Fluids}\ \textbf {\bibinfo {volume} {6}},\ \bibinfo {pages} {064201}} (\bibinfo {year} {2021})}\BibitemShut
  {NoStop}%
\bibitem [{\citenamefont {Taylor}(1953)}]{Taylor:PRSA1953}%
  \BibitemOpen
  \bibfield  {author} {\bibinfo {author} {\bibfnamefont {G.~I.}\ \bibnamefont
  {Taylor}},\ }\bibfield  {title} {\bibinfo {title} {Dispersion of soluble
  matter in solvent flowing slowly through a tube},\ }\href
  {https://doi.org/10.1098/rspa.1953.0139} {\bibfield  {journal} {\bibinfo
  {journal} {Proceedings of the Royal Society of London. Series A. Mathematical
  and Physical Sciences}\ }\textbf {\bibinfo {volume} {219}},\ \bibinfo {pages}
  {186} (\bibinfo {year} {1953})}\BibitemShut {NoStop}%
\bibitem [{\citenamefont {Aris}(1956)}]{Aris:PRSA1956}%
  \BibitemOpen
  \bibfield  {author} {\bibinfo {author} {\bibfnamefont {R.}~\bibnamefont
  {Aris}},\ }\bibfield  {title} {\bibinfo {title} {On the dispersion of a
  solute in a fluid flowing through a tube},\ }\href
  {https://doi.org/10.1098/rspa.1956.0065} {\bibfield  {journal} {\bibinfo
  {journal} {Proceedings of the Royal Society of London. Series A. Mathematical
  and Physical Sciences}\ }\textbf {\bibinfo {volume} {235}},\ \bibinfo {pages}
  {67} (\bibinfo {year} {1956})}\BibitemShut {NoStop}%
\bibitem [{\citenamefont {Chakrabarti}\ and\ \citenamefont
  {Saintillan}(2020)}]{Chakrabarti:PoF2020}%
  \BibitemOpen
  \bibfield  {author} {\bibinfo {author} {\bibfnamefont {B.}~\bibnamefont
  {Chakrabarti}}\ and\ \bibinfo {author} {\bibfnamefont {D.}~\bibnamefont
  {Saintillan}},\ }\bibfield  {title} {\bibinfo {title} {Shear-induced
  dispersion in peristaltic flow},\ }\href {https://doi.org/10.1063/5.0030569}
  {\bibfield  {journal} {\bibinfo  {journal} {Physics of Fluids}\ }\textbf
  {\bibinfo {volume} {32}},\ \bibinfo {pages} {113102} (\bibinfo {year}
  {2020})}\BibitemShut {NoStop}%
\bibitem [{\citenamefont {Darhuber}\ \emph {et~al.}(2004)\citenamefont
  {Darhuber}, \citenamefont {Chen}, \citenamefont {Davis},\ and\ \citenamefont
  {Troian}}]{Darhuber:PRSA2004}%
  \BibitemOpen
  \bibfield  {author} {\bibinfo {author} {\bibfnamefont {A.~A.}\ \bibnamefont
  {Darhuber}}, \bibinfo {author} {\bibfnamefont {J.~Z.}\ \bibnamefont {Chen}},
  \bibinfo {author} {\bibfnamefont {J.~M.}\ \bibnamefont {Davis}},\ and\
  \bibinfo {author} {\bibfnamefont {S.~M.}\ \bibnamefont {Troian}},\ }\bibfield
   {title} {\bibinfo {title} {A study of mixing in thermocapillary flows on
  micropatterned surfaces},\ }\href {https://doi.org/10.1098/rsta.2003.1361}
  {\bibfield  {journal} {\bibinfo  {journal} {Philosophical Transactions of the
  Royal Society of London. Series A: Mathematical, Physical and Engineering
  Sciences}\ }\textbf {\bibinfo {volume} {362}},\ \bibinfo {pages} {1037}
  (\bibinfo {year} {2004})}\BibitemShut {NoStop}%
\bibitem [{sup()}]{supplement}%
  \BibitemOpen
  \href@noop {} {}\bibinfo {note} {See supplemental material at
  http://link.aps.org/supplemental/... for additional details about the
  experiments and the simulations, which includes Refs.~[48, 49]}\BibitemShut
  {NoStop}%
\bibitem [{\citenamefont {Leonard}\ and\ \citenamefont
  {Mokhtari}(1990)}]{Leonard:IJNME1990}%
  \BibitemOpen
  \bibfield  {author} {\bibinfo {author} {\bibfnamefont {B.~P.}\ \bibnamefont
  {Leonard}}\ and\ \bibinfo {author} {\bibfnamefont {S.}~\bibnamefont
  {Mokhtari}},\ }\bibfield  {title} {\bibinfo {title} {Beyond first-order
  upwinding: The ultra-sharp alternative for non-oscillatory steady-state
  simulation of convection},\ }\href {https://doi.org/10.1002/nme.1620300412}
  {\bibfield  {journal} {\bibinfo  {journal} {International Journal for
  Numerical Methods in Engineering}\ }\textbf {\bibinfo {volume} {30}},\
  \bibinfo {pages} {729} (\bibinfo {year} {1990})}\BibitemShut {NoStop}%
\bibitem [{\citenamefont {Cohen}\ \emph {et~al.}(1996)\citenamefont {Cohen},
  \citenamefont {Hindmarsh},\ and\ \citenamefont {Dubois}}]{Cohen:CP1996}%
  \BibitemOpen
  \bibfield  {author} {\bibinfo {author} {\bibfnamefont {S.~D.}\ \bibnamefont
  {Cohen}}, \bibinfo {author} {\bibfnamefont {A.~C.}\ \bibnamefont
  {Hindmarsh}},\ and\ \bibinfo {author} {\bibfnamefont {P.~F.}\ \bibnamefont
  {Dubois}},\ }\bibfield  {title} {\bibinfo {title} {{CVODE}, a stiff/nonstiff
  {ODE} solver in c},\ }\href {https://doi.org/10.1063/1.4822377} {\bibfield
  {journal} {\bibinfo  {journal} {Computers in Physics}\ }\textbf {\bibinfo
  {volume} {10}},\ \bibinfo {pages} {138} (\bibinfo {year} {1996})}\BibitemShut
  {NoStop}%
\bibitem [{\citenamefont {Moosavi}\ and\ \citenamefont
  {Rostami}(2017)}]{Moosavi:JCED2017}%
  \BibitemOpen
  \bibfield  {author} {\bibinfo {author} {\bibfnamefont {M.}~\bibnamefont
  {Moosavi}}\ and\ \bibinfo {author} {\bibfnamefont {A.~A.}\ \bibnamefont
  {Rostami}},\ }\bibfield  {title} {\bibinfo {title} {Densities, viscosities,
  refractive indices, and excess properties of aqueous 1,2-etanediol,
  1,3-propanediol, 1,4-butanediol, and 1,5-pentanediol binary mixtures},\
  }\href {https://doi.org/10.1021/acs.jced.6b00526} {\bibfield  {journal}
  {\bibinfo  {journal} {Journal of Chemical and Engineering Data}\ }\textbf
  {\bibinfo {volume} {62}},\ \bibinfo {pages} {156} (\bibinfo {year}
  {2017})}\BibitemShut {NoStop}%
\bibitem [{\citenamefont {George}\ and\ \citenamefont
  {Sastry}(2003)}]{George:JCED2003}%
  \BibitemOpen
  \bibfield  {author} {\bibinfo {author} {\bibfnamefont {J.}~\bibnamefont
  {George}}\ and\ \bibinfo {author} {\bibfnamefont {N.~V.}\ \bibnamefont
  {Sastry}},\ }\bibfield  {title} {\bibinfo {title} {Density, dynamic
  viscosities, speeds of sound, and relative permittivities for water +
  alkanediols (propane-1,2- and -1,3-diol and butane-1,2-, -1,3-, -1,4-, and
  -2,3-diol) at different temperatures},\ }\href
  {https://doi.org/10.1021/je0340755} {\bibfield  {journal} {\bibinfo
  {journal} {Journal of Chemical and Engineering Data}\ }\textbf {\bibinfo
  {volume} {48}},\ \bibinfo {pages} {1529} (\bibinfo {year}
  {2003})}\BibitemShut {NoStop}%
\bibitem [{\citenamefont {Jarosiewicz}\ \emph {et~al.}(2004)\citenamefont
  {Jarosiewicz}, \citenamefont {Czechowski},\ and\ \citenamefont
  {Jad\.{z}yn}}]{Jarosiewicz:ZNA}%
  \BibitemOpen
  \bibfield  {author} {\bibinfo {author} {\bibfnamefont {P.}~\bibnamefont
  {Jarosiewicz}}, \bibinfo {author} {\bibfnamefont {G.}~\bibnamefont
  {Czechowski}},\ and\ \bibinfo {author} {\bibfnamefont {J.}~\bibnamefont
  {Jad\.{z}yn}},\ }\bibfield  {title} {\bibinfo {title} {The viscous properties
  of diols. v. 1,2-hexanediol in water and butanol solutions},\ }\href
  {https://doi.org/10.1515/zna-2004-0905} {\bibfield  {journal} {\bibinfo
  {journal} {Verlag der Zeitschrift f\"{u}r Naturforschung A}\ }\textbf
  {\bibinfo {volume} {59}},\ \bibinfo {pages} {559} (\bibinfo {year}
  {2004})}\BibitemShut {NoStop}%
\bibitem [{\citenamefont {Eggers}\ and\ \citenamefont
  {Pismen}(2010)}]{Eggers:PoF2010}%
  \BibitemOpen
  \bibfield  {author} {\bibinfo {author} {\bibfnamefont {J.}~\bibnamefont
  {Eggers}}\ and\ \bibinfo {author} {\bibfnamefont {L.~M.}\ \bibnamefont
  {Pismen}},\ }\bibfield  {title} {\bibinfo {title} {Nonlocal description of
  evaporating drops},\ }\href {https://doi.org/10.1063/1.3491133} {\bibfield
  {journal} {\bibinfo  {journal} {Physics of Fluids}\ }\textbf {\bibinfo
  {volume} {22}},\ \bibinfo {pages} {112101} (\bibinfo {year}
  {2010})}\BibitemShut {NoStop}%
\bibitem [{\citenamefont {Verdaguer}\ \emph {et~al.}(2007)\citenamefont
  {Verdaguer}, \citenamefont {Weis}, \citenamefont {Oncins}, \citenamefont
  {Ketteler}, \citenamefont {Bluhm},\ and\ \citenamefont
  {Salmeron}}]{Verdaguer:L2007}%
  \BibitemOpen
  \bibfield  {author} {\bibinfo {author} {\bibfnamefont {A.}~\bibnamefont
  {Verdaguer}}, \bibinfo {author} {\bibfnamefont {C.}~\bibnamefont {Weis}},
  \bibinfo {author} {\bibfnamefont {G.}~\bibnamefont {Oncins}}, \bibinfo
  {author} {\bibfnamefont {G.}~\bibnamefont {Ketteler}}, \bibinfo {author}
  {\bibfnamefont {H.}~\bibnamefont {Bluhm}},\ and\ \bibinfo {author}
  {\bibfnamefont {M.}~\bibnamefont {Salmeron}},\ }\bibfield  {title} {\bibinfo
  {title} {Growth and structure of water on {SiO}2 films on si investigated by
  kelvin probe microscopy and in situ x-ray spectroscopies},\ }\href
  {https://doi.org/10.1021/la700893w} {\bibfield  {journal} {\bibinfo
  {journal} {Langmuir}\ }\textbf {\bibinfo {volume} {23}},\ \bibinfo {pages}
  {9699} (\bibinfo {year} {2007})}\BibitemShut {NoStop}%
\bibitem [{\citenamefont {Barnette}\ \emph {et~al.}(2008)\citenamefont
  {Barnette}, \citenamefont {Asay},\ and\ \citenamefont
  {Kim}}]{Barnette:PCCP2008}%
  \BibitemOpen
  \bibfield  {author} {\bibinfo {author} {\bibfnamefont {A.~L.}\ \bibnamefont
  {Barnette}}, \bibinfo {author} {\bibfnamefont {D.~B.}\ \bibnamefont {Asay}},\
  and\ \bibinfo {author} {\bibfnamefont {S.~H.}\ \bibnamefont {Kim}},\
  }\bibfield  {title} {\bibinfo {title} {Average molecular orientations in the
  adsorbed water layers on silicon oxide in ambient conditions},\ }\href
  {https://doi.org/10.1039/b810309g} {\bibfield  {journal} {\bibinfo  {journal}
  {Physical Chemistry Chemical Physics}\ }\textbf {\bibinfo {volume} {10}},\
  \bibinfo {pages} {4981} (\bibinfo {year} {2008})}\BibitemShut {NoStop}%
\bibitem [{\citenamefont {Matar}(2002)}]{Matar:PoF2002}%
  \BibitemOpen
  \bibfield  {author} {\bibinfo {author} {\bibfnamefont {O.~K.}\ \bibnamefont
  {Matar}},\ }\bibfield  {title} {\bibinfo {title} {Nonlinear evolution of thin
  free viscous films in the presence of soluble surfactant},\ }\href
  {https://doi.org/10.1063/1.1516597} {\bibfield  {journal} {\bibinfo
  {journal} {Physics of Fluids}\ }\textbf {\bibinfo {volume} {14}},\ \bibinfo
  {pages} {4216} (\bibinfo {year} {2002})}\BibitemShut {NoStop}%
\bibitem [{\citenamefont {Karpitschka}\ and\ \citenamefont
  {Riegler}(2010)}]{Karpitschka:L2010}%
  \BibitemOpen
  \bibfield  {author} {\bibinfo {author} {\bibfnamefont {S.}~\bibnamefont
  {Karpitschka}}\ and\ \bibinfo {author} {\bibfnamefont {H.}~\bibnamefont
  {Riegler}},\ }\bibfield  {title} {\bibinfo {title} {Quantitative experimental
  study on the transition between fast and delayed coalescence of sessile
  droplets with different but completely miscible liquids},\ }\href
  {https://doi.org/10.1021/la1007457} {\bibfield  {journal} {\bibinfo
  {journal} {Langmuir}\ }\textbf {\bibinfo {volume} {26}},\ \bibinfo {pages}
  {11823} (\bibinfo {year} {2010})}\BibitemShut {NoStop}%
\bibitem [{\citenamefont {Karpitschka}\ \emph {et~al.}(2015)\citenamefont
  {Karpitschka}, \citenamefont {Weber},\ and\ \citenamefont
  {Riegler}}]{Karpitschka:CES2015}%
  \BibitemOpen
  \bibfield  {author} {\bibinfo {author} {\bibfnamefont {S.}~\bibnamefont
  {Karpitschka}}, \bibinfo {author} {\bibfnamefont {C.}~\bibnamefont {Weber}},\
  and\ \bibinfo {author} {\bibfnamefont {H.}~\bibnamefont {Riegler}},\
  }\bibfield  {title} {\bibinfo {title} {Spin casting of dilute solutions:
  Vertical composition profile during hydrodynamic-evaporative film thinning},\
  }\href {https://doi.org/10.1016/j.ces.2015.01.028} {\bibfield  {journal}
  {\bibinfo  {journal} {Chem. Eng. Sci.}\ }\textbf {\bibinfo {volume} {129}},\
  \bibinfo {pages} {243} (\bibinfo {year} {2015})}\BibitemShut {NoStop}%
\bibitem [{\citenamefont {Hennessy}\ \emph {et~al.}(2017)\citenamefont
  {Hennessy}, \citenamefont {Ferretti}, \citenamefont {Cabral},\ and\
  \citenamefont {Matar}}]{Hennessy:JCIS2017}%
  \BibitemOpen
  \bibfield  {author} {\bibinfo {author} {\bibfnamefont {M.~G.}\ \bibnamefont
  {Hennessy}}, \bibinfo {author} {\bibfnamefont {G.~L.}\ \bibnamefont
  {Ferretti}}, \bibinfo {author} {\bibfnamefont {J.~T.}\ \bibnamefont
  {Cabral}},\ and\ \bibinfo {author} {\bibfnamefont {O.~K.}\ \bibnamefont
  {Matar}},\ }\bibfield  {title} {\bibinfo {title} {A minimal model for solvent
  evaporation and absorption in thin films},\ }\href
  {https://doi.org/10.1016/j.jcis.2016.10.074} {\bibfield  {journal} {\bibinfo
  {journal} {Journal of Colloid and Interface Science}\ }\textbf {\bibinfo
  {volume} {488}},\ \bibinfo {pages} {61} (\bibinfo {year} {2017})}\BibitemShut
  {NoStop}%
\bibitem [{\citenamefont {Oron}\ and\ \citenamefont
  {Nepomnyashchy}(2004)}]{Oron:PRE2004}%
  \BibitemOpen
  \bibfield  {author} {\bibinfo {author} {\bibfnamefont {A.}~\bibnamefont
  {Oron}}\ and\ \bibinfo {author} {\bibfnamefont {A.~A.}\ \bibnamefont
  {Nepomnyashchy}},\ }\bibfield  {title} {\bibinfo {title} {Long-wavelength
  thermocapillary instability with the soret effect},\ }\href
  {https://doi.org/10.1103/physreve.69.016313} {\bibfield  {journal} {\bibinfo
  {journal} {Physical Review E}\ }\textbf {\bibinfo {volume} {69}},\ \bibinfo
  {pages} {016313} (\bibinfo {year} {2004})}\BibitemShut {NoStop}%
\bibitem [{\citenamefont {Shklyaev}\ \emph {et~al.}(2007)\citenamefont
  {Shklyaev}, \citenamefont {Nepomnyashchy},\ and\ \citenamefont
  {Oron}}]{Shklyaev:PoF2007}%
  \BibitemOpen
  \bibfield  {author} {\bibinfo {author} {\bibfnamefont {S.}~\bibnamefont
  {Shklyaev}}, \bibinfo {author} {\bibfnamefont {A.~A.}\ \bibnamefont
  {Nepomnyashchy}},\ and\ \bibinfo {author} {\bibfnamefont {A.}~\bibnamefont
  {Oron}},\ }\bibfield  {title} {\bibinfo {title} {Three-dimensional
  oscillatory long-wave marangoni convection in a binary liquid layer with the
  soret effect: Bifurcation analysis},\ }\href
  {https://doi.org/10.1063/1.2749305} {\bibfield  {journal} {\bibinfo
  {journal} {Physics of Fluids}\ }\textbf {\bibinfo {volume} {19}},\ \bibinfo
  {pages} {072105} (\bibinfo {year} {2007})}\BibitemShut {NoStop}%
\bibitem [{\citenamefont {Shklyaev}\ \emph {et~al.}(2011)\citenamefont
  {Shklyaev}, \citenamefont {Nepomnyashchy},\ and\ \citenamefont
  {Oron}}]{Shklyaev:PRE2011}%
  \BibitemOpen
  \bibfield  {author} {\bibinfo {author} {\bibfnamefont {S.}~\bibnamefont
  {Shklyaev}}, \bibinfo {author} {\bibfnamefont {A.~A.}\ \bibnamefont
  {Nepomnyashchy}},\ and\ \bibinfo {author} {\bibfnamefont {A.}~\bibnamefont
  {Oron}},\ }\bibfield  {title} {\bibinfo {title} {Oscillatory long-wave
  marangoni convection in a layer of a binary liquid: Hexagonal patterns},\
  }\href {https://doi.org/10.1103/physreve.84.056327} {\bibfield  {journal}
  {\bibinfo  {journal} {Physical Review E}\ }\textbf {\bibinfo {volume} {84}},\
  \bibinfo {pages} {056327} (\bibinfo {year} {2011})}\BibitemShut {NoStop}%
\bibitem [{\citenamefont {Thiele}(2011)}]{Thiele:EPJST2011}%
  \BibitemOpen
  \bibfield  {author} {\bibinfo {author} {\bibfnamefont {U.}~\bibnamefont
  {Thiele}},\ }\bibfield  {title} {\bibinfo {title} {Note on thin film
  equations for solutions and suspensions},\ }\href
  {https://doi.org/10.1140/epjst/e2011-01462-7} {\bibfield  {journal} {\bibinfo
   {journal} {The European Physical Journal Special Topics}\ }\textbf {\bibinfo
  {volume} {197}},\ \bibinfo {pages} {213} (\bibinfo {year}
  {2011})}\BibitemShut {NoStop}%
\bibitem [{\citenamefont {Thiele}\ \emph {et~al.}(2013)\citenamefont {Thiele},
  \citenamefont {Todorova},\ and\ \citenamefont {Lopez}}]{Thiele:PRL2013}%
  \BibitemOpen
  \bibfield  {author} {\bibinfo {author} {\bibfnamefont {U.}~\bibnamefont
  {Thiele}}, \bibinfo {author} {\bibfnamefont {D.~V.}\ \bibnamefont
  {Todorova}},\ and\ \bibinfo {author} {\bibfnamefont {H.}~\bibnamefont
  {Lopez}},\ }\bibfield  {title} {\bibinfo {title} {Gradient dynamics
  description for films of mixtures and suspensions: Dewetting triggered by
  coupled film height and concentration fluctuations},\ }\href
  {https://doi.org/10.1103/physrevlett.111.117801} {\bibfield  {journal}
  {\bibinfo  {journal} {Physical Review Letters}\ }\textbf {\bibinfo {volume}
  {111}},\ \bibinfo {pages} {117801} (\bibinfo {year} {2013})}\BibitemShut
  {NoStop}%
\bibitem [{\citenamefont {Jensen}\ and\ \citenamefont
  {Grotberg}(1993)}]{Jensen:PoFA1993}%
  \BibitemOpen
  \bibfield  {author} {\bibinfo {author} {\bibfnamefont {O.~E.}\ \bibnamefont
  {Jensen}}\ and\ \bibinfo {author} {\bibfnamefont {J.~B.}\ \bibnamefont
  {Grotberg}},\ }\bibfield  {title} {\bibinfo {title} {The spreading of heat or
  soluble surfactant along a thin liquid film},\ }\href
  {https://doi.org/10.1063/1.858789} {\bibfield  {journal} {\bibinfo  {journal}
  {Physics of Fluids A: Fluid Dynamics}\ }\textbf {\bibinfo {volume} {5}},\
  \bibinfo {pages} {58} (\bibinfo {year} {1993})}\BibitemShut {NoStop}%
\bibitem [{\citenamefont {Jensen}\ \emph {et~al.}(1994)\citenamefont {Jensen},
  \citenamefont {Halpern},\ and\ \citenamefont {Grotberg}}]{Jensen:CES1994}%
  \BibitemOpen
  \bibfield  {author} {\bibinfo {author} {\bibfnamefont {O.~E.}\ \bibnamefont
  {Jensen}}, \bibinfo {author} {\bibfnamefont {D.}~\bibnamefont {Halpern}},\
  and\ \bibinfo {author} {\bibfnamefont {J.~B.}\ \bibnamefont {Grotberg}},\
  }\bibfield  {title} {\bibinfo {title} {Transport of a passive solute by
  surfactant-driven flows},\ }\href
  {https://doi.org/10.1016/0009-2509(94)85083-6} {\bibfield  {journal}
  {\bibinfo  {journal} {Chemical Engineering Science}\ }\textbf {\bibinfo
  {volume} {49}},\ \bibinfo {pages} {1107} (\bibinfo {year}
  {1994})}\BibitemShut {NoStop}%
\bibitem [{\citenamefont {Diez}\ \emph {et~al.}(2000)\citenamefont {Diez},
  \citenamefont {Kondic},\ and\ \citenamefont {Bertozzi}}]{Diez:PRE2000}%
  \BibitemOpen
  \bibfield  {author} {\bibinfo {author} {\bibfnamefont {J.~A.}\ \bibnamefont
  {Diez}}, \bibinfo {author} {\bibfnamefont {L.}~\bibnamefont {Kondic}},\ and\
  \bibinfo {author} {\bibfnamefont {A.}~\bibnamefont {Bertozzi}},\ }\bibfield
  {title} {\bibinfo {title} {Global models for moving contact lines},\ }\href
  {https://doi.org/10.1103/physreve.63.011208} {\bibfield  {journal} {\bibinfo
  {journal} {Physical Review E}\ }\textbf {\bibinfo {volume} {63}},\ \bibinfo
  {pages} {011208} (\bibinfo {year} {2000})}\BibitemShut {NoStop}%
\bibitem [{\citenamefont {Lenz}\ \emph {et~al.}(2002)\citenamefont {Lenz},
  \citenamefont {Rumpf},\ and\ \citenamefont {Gr\"un}}]{Gruen2002}%
  \BibitemOpen
  \bibfield  {author} {\bibinfo {author} {\bibfnamefont {M.}~\bibnamefont
  {Lenz}}, \bibinfo {author} {\bibfnamefont {M.}~\bibnamefont {Rumpf}},\ and\
  \bibinfo {author} {\bibfnamefont {G.}~\bibnamefont {Gr\"un}},\ }\bibfield
  {title} {\bibinfo {title} {A finite volume scheme for surfactant driven thin
  film flow},\ }in\ \href@noop {} {\emph {\bibinfo {booktitle} {Proceedings of
  the Third International Symposium on Finite Volumes for Complex
  Applications}}},\ \bibinfo {editor} {edited by\ \bibinfo {editor}
  {\bibfnamefont {R.}~\bibnamefont {Herbin}}\ and\ \bibinfo {editor}
  {\bibfnamefont {D.}~\bibnamefont {Kr\"oner}}}\ (\bibinfo  {publisher} {Hermes
  Penton Science},\ \bibinfo {year} {2002})\ pp.\ \bibinfo {pages}
  {567--574}\BibitemShut {NoStop}%
\bibitem [{\citenamefont {Charlier}\ \emph {et~al.}(2022)\citenamefont
  {Charlier}, \citenamefont {Rednikov}, \citenamefont {Dehaeck}, \citenamefont
  {Colinet},\ and\ \citenamefont {Terwagne}}]{Charlier:JFM2022}%
  \BibitemOpen
  \bibfield  {author} {\bibinfo {author} {\bibfnamefont {J.}~\bibnamefont
  {Charlier}}, \bibinfo {author} {\bibfnamefont {A.}~\bibnamefont {Rednikov}},
  \bibinfo {author} {\bibfnamefont {S.}~\bibnamefont {Dehaeck}}, \bibinfo
  {author} {\bibfnamefont {P.}~\bibnamefont {Colinet}},\ and\ \bibinfo {author}
  {\bibfnamefont {D.}~\bibnamefont {Terwagne}},\ }\bibfield  {title} {\bibinfo
  {title} {Water-propylene glycol sessile droplet shapes and migration:
  Marangoni mixing and separation of scales},\ }\href
  {https://doi.org/10.1017/jfm.2021.1030} {\bibfield  {journal} {\bibinfo
  {journal} {J. Fluid Mech.}\ }\textbf {\bibinfo {volume} {933}},\ \bibinfo
  {pages} {A45} (\bibinfo {year} {2022})}\BibitemShut {NoStop}%
\end{thebibliography}

%

\end{document}